\documentclass[12pt]{article}
\usepackage{eurosym}
\usepackage{graphicx}
\usepackage[utf8]{inputenc}
\usepackage[T1]{fontenc}
\usepackage{indentfirst}
\usepackage[margin=0.6in,nomarginpar]{geometry}
\usepackage[final]{hyperref}
\usepackage{amsmath}
\usepackage{hyperref}
\usepackage{cite}
\usepackage{subcaption}
\usepackage{caption}
\usepackage{amssymb}
\usepackage{multirow}
\usepackage[table]{xcolor}

\hypersetup{
colorlinks=true,
linkcolor=blue,
citecolor=blue,
filecolor=magenta,
urlcolor=blue
}

\begin{document}

\title{Noncommutative Schwarzschild black hole surrounded by quintessence:
Thermodynamics, Shadows and Quasinormal modes}
\author{B. Hamil\thanks{%
hamilbilel@gmail.com, bilel.hamil@edu.umc.dz} \\
Laboratoire de Physique Math\'{e}matique et Subatomique,\\
Facult\'{e} des Sciences Exactes, Universit\'{e} Constantine 1, Constantine,
Algeria. \and B. C. L\"{u}tf\"{u}o\u{g}lu\thanks{%
bekir.lutfuoglu@uhk.cz (Corresponding author)} \\
Department of Physics, University of Hradec Kr\'{a}lov\'{e},\\
Rokitansk\'{e}ho 62, 500 03 Hradec Kr\'{a}lov\'{e}, Czechia.}
\date{\today }
\maketitle

\begin{abstract}
{In (Sci. Rep. \textbf{12}, 8516 (2022)), Campos et al studied}
the quasinormal modes and shadows of noncommutative Schwarzschild black
holes. Since we know that the quintessence matter surrounding black holes
has { significant effects} on the black hole quantities, {%
in this study, we aimed to show this influence by revisiting the
same problem in the presence of quintessence matter field. To this end, we
first examined the thermodynamics of noncommutative Schwarzschild black
holes embedded in quintessence matter using Hawking temperature, entropy,
and specific heat functions. After that,} we discussed phase transition and
stability { features. We then investigated} the shadow images in
the presence of plasma. After visualizing these results qualitatively, we
calculated the quasinormal modes in WKB and Mashhoon approximations and we
demonstrated the impacts of quintessence matter and noncommutative spacetime
on the whole quantities.
\end{abstract}


\section{Introduction}

Before the nineteen seventies, there was no significant motivation to debate
the black holes in the framework of their thermodynamics. This perspective
changed drastically in 1973 with the remarkable paper of Bekenstein \cite%
{Bekenstein}, in which the entropy of a black hole was related to the black
hole's area via Hawking's theorem \cite{Hawking}. In the same year, Bardeen
et al. employed an analogy between surface gravity and temperature in
addition to the existing one between the entropy and event horizon surface
area, so they stated four fundamental laws to investigate black hole
thermodynamics \cite{Bardeen}. Two years later, Hawking refuted the
contradictory ideas of the classical approach that black holes should absorb
all matter and energy without emitting radiation theoretically \cite%
{WHawking}. According to him, quantum effects had to play a critical role in
black holes, thus, black holes could also emit radiation characterized by a
spectrum similar to that of a black body with a specific temperature. In the
following years, Hawking's interpretation was confirmed by many other
independent ways \cite{Carlip}, which led to an increase in studies
discussing the thermal properties of black holes \cite{page2005, Belhaj2012,
Appels2016, Appels2017, HH2019, Andre2020, Musmarra2021, Hamil2022, Wu2022,
Hamil20221, Khosravipoor2023, Wu2023, Sadeghi2024}.

Distinguished observations in the last decade of the last century, based on
the magnitude-redshift relation of astronomical objects, have revealed that
the universe is expanding at an accelerating rate \cite{Riess1, Riess2,
Perlmutter1999}. Since the theoretically predicted value of the cosmological
constant, which is expected to correspond to this phenomenon in the general
theory of relativity via the action, 
\begin{eqnarray}
\mathcal{S}&=&\frac{1}{16 \pi G}\int d^4x\sqrt{-g} \Big[R+2 \Lambda\Big]+%
\mathcal{S_M},
\end{eqnarray}
{ differs from the observational value by $120$ orders of
magnitude \cite{Carroll},} alternative explanations have begun to be
considered. Some theorists thought that dark energy, distributed relatively
uniformly in space with a negative pressure, could be responsible for this 
\cite{Peebles}. In the literature, dark energy is modeled by using dynamic
scalar fields with different state parameter equations \cite{carroll1998,
Khoury2004, Picon2000, Padman2002, Caldwell2002, Gasperini2002, Copeland2006}%
. One of the most examined forms is given by the quintessence matter model
with the action {%
\begin{equation}
\mathcal{S}=\int d^{4}x\sqrt{-g}\bigg[\frac{R}{16\pi G}+\mathcal{L}_{quin}%
\bigg]+\mathcal{S_{M}},  \label{acton11}
\end{equation}%
where the quintessence term is coupled to the action by the Lagrange density
term 
\begin{equation}
\mathcal{L}_{quin}=-\frac{1}{2}\left( \nabla \phi \right) ^{2}-V\left( \phi
\right),
\end{equation}%
with the quintessential scalar field, $\phi $, and potential, $V\left( \phi
\right)$. 

In this theory, the universe's dark energy is primarily dominated by the
potential of the scalar field, which is continuously evolving towards its
minimum at $V=0$. Usually, this minimum is situated at $\phi =\infty $, and
the scalar potential may have a form such as \ $V\sim e^{-c\phi }$. The
theory can also be parameterized by an equation of state of the usual form 
\cite{Yoo, Simeon}%
\begin{equation}
P=\omega _{q}\rho ,
\end{equation}%
where $P$ is the pressure and $\rho $ is the energy density. In this
approach, the equation of state parameter has to be in the range $-1<\omega
_{q}<-1/3$ \cite{Chiba}.} In 2003, Kiselev derived a general form of the
static spherically symmetric solutions of Einstein's equations and presented
the line elements of electrically uncharged and charged black holes
surrounded by quintessence matter \cite{Kiselev}. Following this work, some
other physicists obtained new solutions corresponding to various black holes
of different properties in the presence of quintessence matter \cite%
{Chen2005, Zhang2006, Chen2008, Wei2011, Thomas2012, Fernando2013,
Ghosh2016, XuKerr, Fontana2018, Wu2018, Nozari2020, Nozari2022, Wang2023}.
Recently, with the growing interest in black hole thermodynamics, we have
observed { a growing interest in the} studies discussing the
influence of quintessence matter on the thermal quantities of black holes 
\cite{Ghaderi2016, Ma2017, Ghaderi2018, Liu2019, Shahjalal2019, Haldar2020,
BB1, Ndongmo, Chen, BB2, Zhang, BB3, BB4}.

Until recent years, one of the main concerns about black holes was about
their observations. Even if they could not be observed directly, could their
effects, that is, their fingerprints, be detected? One of the ideas in the
ongoing debates in this direction was based on the argument that real black
holes should not be an ideally isolated system. Accordingly, after the
gravitational collapse of matter, black holes should be formed in a
perturbed state, and thus their fundamental parameters, namely their masses,
charges, and angular momenta should not be enough to discuss their features,
i.e. stability, Hawking radiation... A perturbed black hole is assumed to
oscillate the background by emitting gravitational waves, which are damped
after a short while of the initial outburst of radiation over time. In the
literature, these particular oscillation frequencies are called the
quasinormal modes (QNMs) \cite{Press1971}. Mathematically, these modes
appear in complex number forms, where their real parts denote the actual
oscillation, while their complex parts correspond to the damping time
inversely \cite{Mod2}. Interestingly, QNMs were found to be independent of
the initial perturbation, and thus, they are accepted as the fingerprints of
black holes \cite{Konoplya2011}. Following the great success of the
LIGO-VIRGO collaboration in detecting gravitational waves \cite{BF1, BF2},
interest in QNMs has increased enormously {\cite{Zhidenko,
Rincon2018, Chirenti2018, Tattersall, Myung2019, Kon2019, Salcedo2019,
Kon2020, Liu2020, Jusufi2020, Roman2020, HendiJamil2020, HendiJamil2021,
Kanzi2021, Anacleto2021, Gulnihal2021, Jafarzade, Ghosh2021, Okyay2022,
Ovgun2022, Chen2022, Sakalli2022, Roman2023, Nesrin2023, Anacleto2023,
Jha2023, Lambiase2023, Demir2023, Badawi2024, Das2024}.}

One of the other fingerprints of black holes is their shadows. Very
recently, { with joint work, Event Horizon Telescope} succeeded
in determining black hole shadows \cite{M871, SagA1}. This observation is
based on detecting the photon deflections radiated by strong light sources
in the background of the black holes \cite{He}. According to the modeling,
when light rays pass near a black hole, only photons with low orbital
angular momentum are trapped and photons with high orbital angular momentum
are deflected. Therefore, depending on the observing position, the observer
observes dark regions in the observational sky \cite{Rev1}. Historically,
the theoretical estimation of the concept of shadow dates back half a
century \cite{Synge66, Bardeenk, Luminet}, but it was only after the
aforementioned observations that it began to be studied intensively {%
 \cite{Tsu1, Tsu2, Tsu3, Konoplyae, WeiMann, Babar, Kumar, Singh,
Nodehi2020, LiGuo, Zhang21, Thomas1, AtamuratovJamil2021, Rayimbaev2022,
Das, Pantig2022, GuoWD, PG1, ZG1, Sunny1, Sunny2, Sunny3, Sunny4, Sunny5,
Sunny6, Vir, Vir1, Biz2023, Biz2023b, Du, Akhil, Kara, Ali1, Chowdhuri,
Molla, Olmo, AtamuratovJamil2023, Hoshimov2024}.}


The final stage of black hole evaporation is still debated in the
literature. According to some views, string effects should be taken into
account at this stage, and noncommutative geometry, which has a long history 
\cite{Snyder1947}, could be a suitable approach to account for these extreme
quantum gravitational effects \cite{Niki2009}. From this point of view, in
2005 two independent studies, first Nicollini \cite{Nicollini2005} and then
Nasseri \cite{Nasseri2005}, considered Schwarzchild black holes in
non-commutative geometry. The following year, Nicolini et al demonstrated
that noncommutative effects vanish several problems in the evaporation
process \cite{Niki2006}. Then, Rizzo showed that in the presence of extra
dimensions significant modifications emerge \cite{Rizzo2006}. After these
cornerstone studies, other black holes and their features have also been
extensively investigated in noncommutative geometry \cite{Myung2007,
Dolan2007, Nozari2007, Kim2008, Banerjee2008, Nozari2008, Arraut2009,
Niko2009, Mehdi2010, Sharif2011, Rahaman2013, Marcos2014, Marcos2015,
Miao2017, Macros2018, Soto2018, Marcos2020, Marcos2021, Marcos2021b,
Crespo2022, Juric2023, Macros2023}. Recently, in \cite{Liang2018,
Liang2018b, Yan2020, Marcos2022, Zhao2023} authors examined the QNMs of the
Schwarzschild-like black holes. Moreover, Campos et al derived the shadow
radius and discussed the impact of the noncommutativity parameter on it in 
\cite{Marcos2022}.

Inspired by all these facts, in this manuscript, we intend to determine the
impact of the quintessence matter on the thermal quantities, stability, and
phase transitions of the Schwarzschild black hole in noncommutative
spacetime. Moreover, we aim to investigate the shadows and QNMs of the black
hole and demonstrate the influence of the quintessence matter on them. To
this end, we construct the manuscript as follows: In Sec. \ref{sec2}, we
present a brief of the noncommutative effects on the black hole mass and
lapse functions. Then, in Sec. \ref{sec3}, we examine the black hole
thermodynamics. Next, in Sec. \ref{sec4}, we obtain the expected shadow
images of the black hole. Then, in Sec. \ref{sec5}, discuss the QNMs.
Finally, we conclude the manuscript.



\section{A brief review}

\label{sec2}

{ In this section, we aim to introduce the geometry of the
noncommutative Schwarzschild black hole that is surrounded by the
quintessence matter field. Here, noncommutativity can be taken as a
correction to the Schwarzschild black hole metric and its contribution
vanishes when its strength goes to zero, as defined in the simplest form
below: 
\begin{equation}
\left[ X^{\mu },X^{\nu }\right] =i\Theta ^{\mu \nu }.
\end{equation}%
Here, $\Theta ^{\mu \nu }$ is an antisymmetric constant tensor of dimension
(length)$^{2}$.

In commutative spacetime, one can represent a point particle's mass density
by an ordinary product of its mass with a Dirac delta function. However, in
a noncommutative space, describing point mass in such a manner becomes
impractical due to the inherent fuzziness of space resulting from the
position-position uncertainty relation. In this case, the measure of this
fuzziness has to be considered with the noncommutative parameter, $\Theta $.
In literature, various forms of mass density have been proposed \cite%
{Nicollini2005, Marcos2022, Giri2007, Farook, Idris}. In this manuscript, we employ the
following distribution form \cite{Marcos2022} 
\begin{equation}
\rho _{matt}\left( \Theta ,r\right) =\frac{M\sqrt{\Theta }}{\pi ^{3/2}\left(
r^{2}+\pi \Theta \right) ^{2}},  \label{den}
\end{equation}%
where $M$ is the total mass defused throughout the region of linear size $%
\sqrt{\Theta}$. Now, in the presence of a quintessence matter field, we look
for a static, spherically symmetric, asymptotically Schwarzschild black hole
solution of the Einstein equations with the energy density defined above. We
consider the spherically symmetric black hole metric of the form 
\begin{eqnarray}
ds^{2} &=&-e^{\nu }dt^{2}+e^{\mu }dr^{2}+r^{2}d\Omega ^{2},  \notag \\
&=&-f\left( r, \Theta, \omega _{q}\right) dt^{2}+\frac{1}{f\left( r, \Theta,
\omega _{q}\right) }dr^{2}+r^{2}d\Omega ^{2},  \label{ansatz}
\end{eqnarray}
with 
\begin{equation}
f\left( r, \Theta, \omega _{q}\right) =1-\frac{2\mathcal{M}\left( r, \Theta,
\omega _{q}\right) }{r},
\end{equation}
where $\mu $ and $\nu $ depend only on $r$ with $\nu =-\mu$, and $\mathcal{M}%
\left( r, \Theta, \omega _{q}\right)$ represents the smear mass. We then
express the Einstein equation with the following form 
\begin{equation}
G_{\alpha \beta }=R_{\alpha \beta }-\frac{1}{2}g_{\alpha \beta }R=2\left( 4
\pi T_{{\alpha \beta}_{matt} }\left( r, \Theta \right) +T_{{\alpha \beta}%
_{quin} }\left( \omega _{q}\right) \right) ,  \label{15}
\end{equation}%
where we assume $G=c=\hbar= k_{B}=1.$ As shown by Kiselev in\cite{Kiselev},
we take the time component of the energy-momentum tensor for quintessence
matter 
\begin{equation}
T_{0 _{quin}}^{0}\left( \omega _{q}\right) =\frac{3\omega _{q}\sigma }{%
2r^{3\omega _{q}+3}},
\end{equation}%
and the stress-energy tensor component 
\begin{equation}
T_{0 _{matt}}^{0}\left( r, \Theta \right) =-\rho _{matt}\left( \Theta
,r\right) ,
\end{equation}%
and employ them in Eq. (\ref{15}). We find 
\begin{equation}
G_{0}^{0}=-8 \pi\rho _{matt}\left( \Theta ,r\right) +\frac{3\omega
_{q}\sigma }{r^{3\omega _{q}+3}}.  \label{stein}
\end{equation}
To solve Eq. (\ref{stein}), we first need to determine the $G_{00}$
component of the Einstein tensor. Using the metric, given in Eq. (\ref%
{ansatz}), we calculate it as: 
\begin{equation}
G_{00}=e^{\nu }\left( \frac{1}{r^{2}}-e^{-\mu }\left( \frac{1}{r^{2}}-\frac{%
\mu ^{\prime }}{r}\right) \right) =f\left( r, \Theta, \omega _{q}\right)
\left( \frac{2}{r^{2}}\frac{d\mathcal{M}\left( r, \Theta, \omega _{q}\right) 
}{dr}\right) .  \label{28}
\end{equation}%
Thus, $G_{0}^{0}$ component reads: 
\begin{equation}
G_{0}^{0}=-\frac{2}{r^{2}}\frac{d\mathcal{M}\left( r, \Theta, \omega
_{q}\right) }{dr}.  \label{sk}
\end{equation}
Then, we match Eqs. (\ref{stein}) and (\ref{sk}), and we get 
\begin{equation}
\frac{d\mathcal{M}\left(r, \Theta, \omega _{q}\right) }{dr}=4\pi r^{2}\rho
_{matt}\left( \Theta ,r\right) -\frac{3\omega _{q}\sigma }{2r^{3\omega
_{q}+1}}.  \label{31}
\end{equation}%
After we substitute the energy distribution, we integrate Eq. (\ref{31}). We
obtain the smeared mass distribution 
\begin{equation}
\mathcal{M}\left( r, \Theta, \omega _{q}\right) =\frac{2M}{\pi }\left[ \tan
^{-1}\left( \frac{r}{\sqrt{\pi \Theta }}\right) -\frac{r\sqrt{\pi \Theta }}{%
r^{2}+\pi \Theta }\right] +\frac{\sigma }{2r^{3\omega _{q}}}.
\label{metric09}
\end{equation}
Therefore, the lapse function reads: 
\begin{equation}
f\left( r, \Theta, \omega _{q}\right) =1-\frac{2M}{\pi }\left[ \tan
^{-1}\left( \frac{r}{\sqrt{\pi \Theta }}\right) -\frac{r\sqrt{\pi \Theta }}{%
r^{2}+\pi \Theta }\right] -\frac{\sigma }{r^{1+3\omega _{q}}},
\label{metric10}
\end{equation}%
and up to the first order of the noncommutative correction it becomes 
\begin{equation}
f\left( r, \Theta, \omega _{q} \right) =1-\frac{2M}{r}+\frac{8M}{\sqrt{\pi }%
r^{2}}\sqrt{\Theta }-\frac{\sigma }{r^{3\omega _{q}+1}}.  \label{metric11}
\end{equation}
It is worth noting that in the absence of the quintessence matter field, the
last terms of Eqs. (\ref{metric09}), (\ref{metric10}) and (\ref{metric11})
drop, so the smeared mass and the lapse functions become the same as given
in \cite{Marcos2020, Marcos2022}. Moreover, for the commutative case, $%
\Theta \rightarrow 0$, Eq. (\ref{metric11}) simplifies to the conventional
Schwarzschild metric surrounded by quintessence \cite{BB1}.

Here, we should also note that for supermassive black holes, like six
million solar mass one, spacetime can be treated as a smooth classical
manifold since the effect of noncommutativity is negligible (commutative
limit). However, for mini black holes \cite{Arraut}, where quantum effects
dramatically alter spacetime structure, substantial differences arise from
spacetime fuzziness related to uncertainty principle considerations. One
expects significant changes due to the spacetime fuzziness. Moreover, for $%
\frac{8M\sqrt{\Theta }}{\sqrt{\pi }}=Q^{2},$ the metric is largely similar
to the Reissner-Nordstr\"om black hole surrounded with quintessence 
\cite{Balendra,Saleh, Peng} }

{ In Figure \ref{fig:met}, we plot the lapse function versus $r$. To be more precise,  in Figures \eqref{fig:meta} and \eqref{fig:metb} we demonstrate the impact of the quintessence field and noncommutativity by comparing the lapse function with the ordinary case one. Then, in Figures \eqref{fig:1c} and \eqref{fig:1d}, we investigate whether a naked singularity exists or not for the two cases that we will use during the rest of the manuscript. 

\begin{figure}[tbh]
\begin{minipage}[t]{0.50\textwidth}
        \centering
  \includegraphics[width=\textwidth]{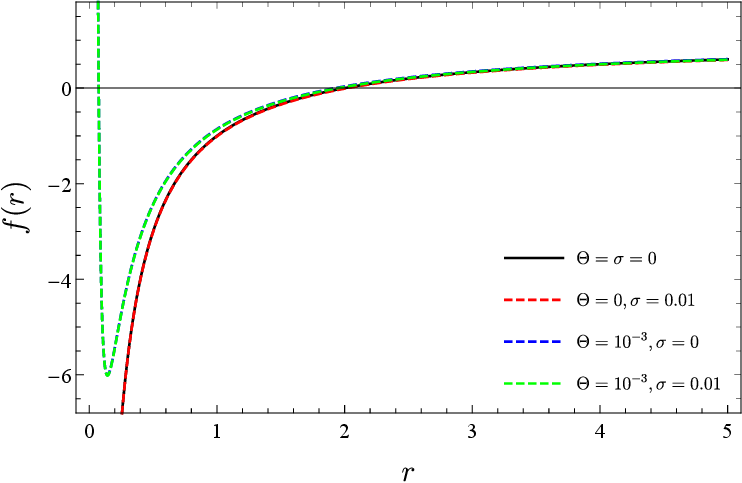}
  \subcaption{ $\omega _{q}=-0.35$} 
            \label{fig:meta}
   \end{minipage}%
\begin{minipage}[t]{0.50\textwidth}
        \centering
       \includegraphics[width=\textwidth]{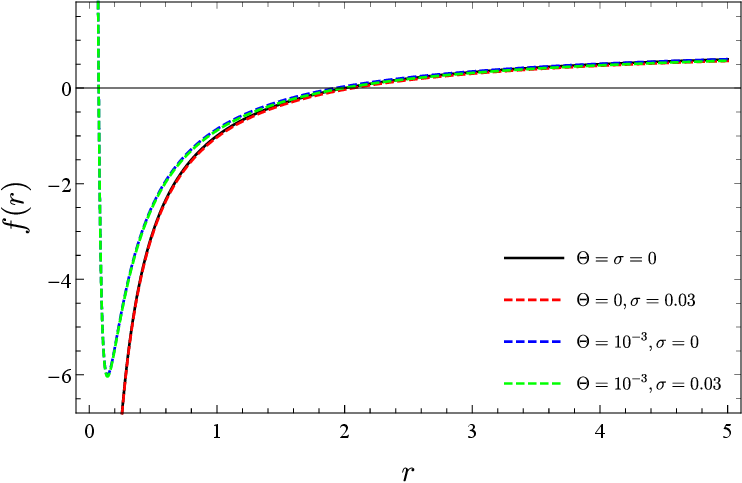}\\
          \subcaption{ $\omega _{q}=-0.35$}   \label{fig:metb}
    \end{minipage}\hfill

\begin{minipage}[t]{0.50\textwidth}
        \centering
       \includegraphics[width=\textwidth]{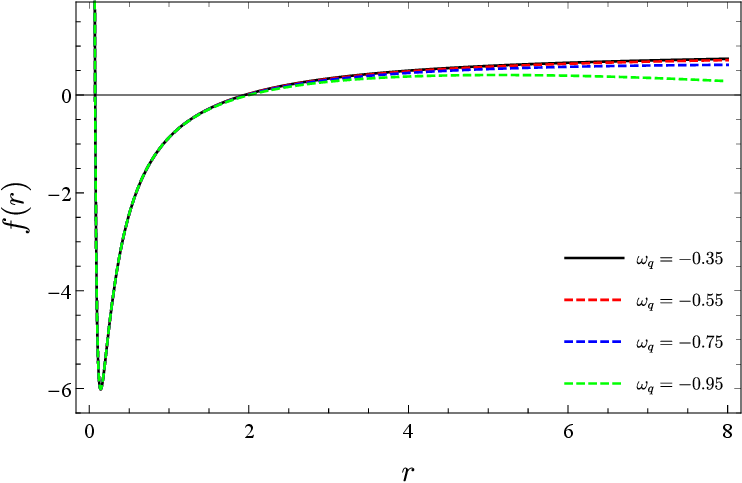}\\
          \subcaption{$\sigma=0.01$}   \label{fig:1c}
    \end{minipage}\hfill
    \begin{minipage}[t]{0.50\textwidth}
        \centering
  \includegraphics[width=\textwidth]{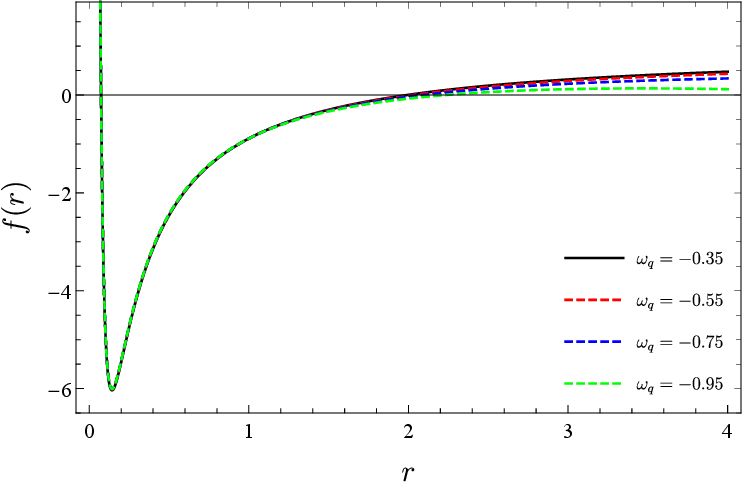}
  \subcaption{$\sigma=0.03$ } 
            \label{fig:1d}
   \end{minipage}%
\caption{Variation of the lapse function with respect to $r$ for $M=1$.}
\label{fig:met}
\end{figure}

\newpage

Here, we have the following observations: 
\begin{itemize}
\item For $\Theta =\sigma =0$, the metric function matches with the standard Schwarzschild solution \cite{Niki2006}.  The well-known solution has a coordinate singularity at $r=2M$ and it is free from naked singularity. 

\item  Without noncommutativity, the quintessence matter-modified metric  function mimics the ordinary case with a slight shift at greater radii values at relatively small radii. For example, for  $\sigma=0.01$ and $\sigma=0.03$ with $\omega_q=-0.35$ the coordinate singularity appears at $r\simeq 2.02M$ and $r\simeq 2.06M$, respectively. In both cases, the event horizons cover the central singularity. Therefore, we conclude that these cases are also free of naked singularity.

\item In the noncommutative case without the quintessence matter the lapse function approaches infinity as $r$ tends to zero. This indicates the presence of a curvature singularity at the origin, $r=0$, however, it is not detectable because of the fuzziness of the spacetime. In addition, we observe that the lapse function changes its sign at $r\simeq 1.93M$. This implies the existence of an event horizon at that radius which shows us that the central singularity is covered by the event horizon.  

\item In the noncommutative case with the quintessence matter the lapse functions mimic the previous case with a slight shift at relatively small radii and it approaches infinity as $r$ tends to zero. For $\sigma=0.01$ and $\sigma=0.03$ with $\omega_q=-0.35$ the lapse functions change their sign at $r\simeq 1.95M$ and $r\simeq 1.99M$, respectively. Like the previous case, we conclude that in this scenario the central singularity is appropriately covered by the event horizon.

\item Figure \eqref{fig:1c} for $\sigma=0.01$ and Figure \eqref{fig:1d} for $\sigma=0.03$ confirm the existence of horizon radii for various quintessence state parameters. 
Since the model with these parameters does not contain any physically unacceptable naked singularities, we will use these parameters  in the remainder of this article.

\end{itemize}
}

\begin{figure}[tbh]
\begin{minipage}[t]{0.5\textwidth}
        \centering
        \includegraphics[width=\textwidth]{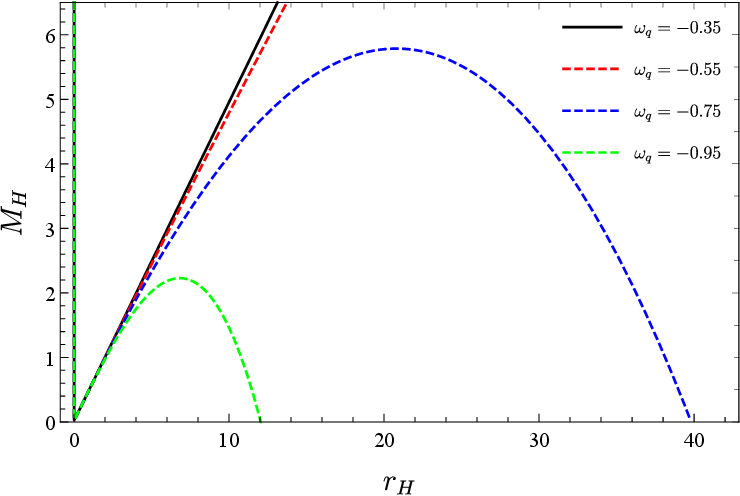}
       \subcaption{ {$\sigma=0.01$} and $\Theta=10^{-4}$.}\label{fig:Ma}
   \end{minipage}%
\begin{minipage}[t]{0.5\textwidth}
        \centering
        \includegraphics[width=\textwidth]{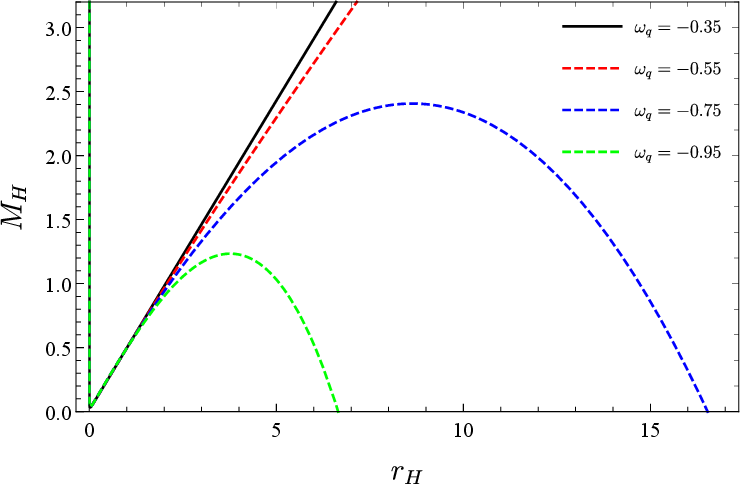}
         \subcaption{ {$\sigma=0.03$} and $\Theta=10^{-4}$.}\label{fig:Mb}
   \end{minipage}\ 
\begin{minipage}[t]{0.5\textwidth}
        \centering
        \includegraphics[width=\textwidth]{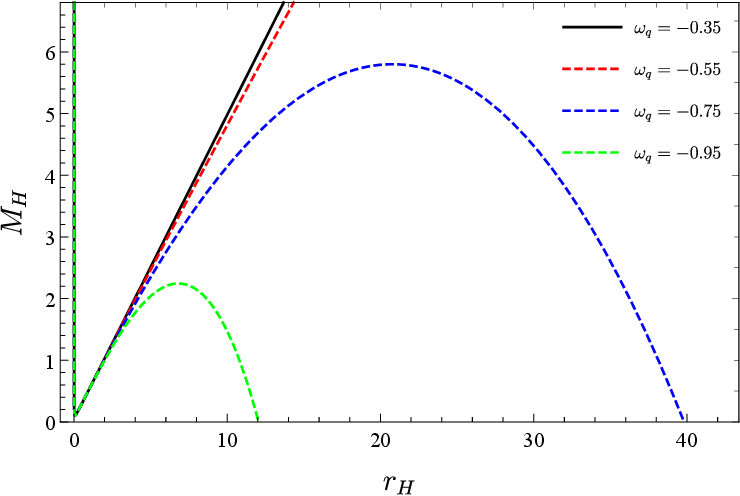}
       \subcaption{ {$ \sigma=0.01$} and $\Theta=10^{-3}$.}\label{fig:Mc}
   \end{minipage}%
\begin{minipage}[t]{0.5\textwidth}
        \centering
        \includegraphics[width=\textwidth]{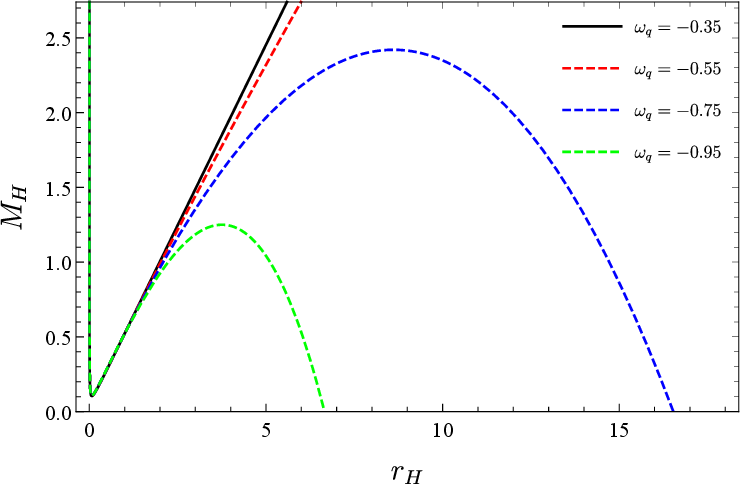}
         \subcaption{ {$\sigma=0.03$} and $\Theta=10^{-3}$.}\label{fig:Md}
   \end{minipage}
\caption{The effect of the quintessence matter on the black hole mass.}
\label{figmass}
\end{figure}

Then, for a specific quintessential state parameter, we determine the event horizon, { $r_{H}$, via $f\left( r_{H},\Theta \right) =0$, and} we express black hole mass function in terms of the event horizon in the presence of quintessence as 
\begin{equation}
M_{H}(\omega _{q},\Theta )={\frac{r_{H}}{2}\left( 1+\frac{4\sqrt{\Theta }}{%
\sqrt{\pi }r_{H}}+\frac{16\Theta }{\pi r_{H}^{2}}\right)} \left( 1-\frac{%
\sigma }{r_{H}^{1+3\omega _{q}}}\right) .  \label{m1}
\end{equation}%
In Figure \ref{figmass}, we depict the mass function for a set of two
different valued noncommutative and normalization constant parameters. In
each plot, we consider four different quintessence state parameters which
correspond to four distinct scenarios. 

These qualitative plots show us that the quintessence matter alters the
characteristics of the mass function as in the commutative case. { As we know well, in the ordinary Schwarzschild black hole case, the mass function has a linear relationship with the horizon radius,   $M=\frac{r_{H}}{2}$. This relation sets forward that as the horizon radius leads to zero, the mass also approaches to zero, thus,  the mass vanishes at the origin. However, in noncommutative spacetime due to fuzziness, there is a lower bound on the radius, which ensures that the black hole mass has a non-zero minimum mass value at a minimum event horizon radius. We observe that at relatively small horizon radii quintessence matter does not have an impact on the mass function. In the limit of  $r_{H}>>\sqrt{\Theta }$, the impact of noncommutative geometry disappears, however, the effects of the quintessence matter become dominant and it enforces a decrease in the rate of the mass increase \cite{Ghaderi2016n}. This alteration ends at a maximum mass value, and then, the mass starts to decrease and it equals to zero at $r_{H}=\sigma ^{\frac{1}{3\omega _{q}+1}}$.  }


\section{Black Hole Thermodynamics}

\label{sec3}

In this section, our focus will be on deriving the exact forms of various
thermal quantities such as Hawking temperature, entropy, heat capacity, and
Gibbs free energy. Our exploration begins with the Hawking temperature,
which is usually defined by the following expression {\cite{WHawking}}
\begin{equation}
T=\frac{\kappa }{2\pi },
\end{equation}%
where $\kappa $ corresponds to the surface gravity and it can be deduced via 
\begin{equation}
\kappa =-\left. \sqrt{-\frac{g^{11}}{g^{00}}}\frac{\left( g_{00}\right)
^{\prime }}{g_{00}}\right\vert _{r=r_{H}}.
\end{equation}%
After performing the straightforward computations, we obtain the Hawking
temperature with the noncommutative correction term in the form of 
\begin{equation}
T_{H}(\omega _{q},\Theta )=\frac{1}{{4}\pi r_{H}}\left( 1+\frac{3\sigma \omega
_{q}}{r_{H}^{3\omega _{q}+1}}\right) -\frac{{1}}{\pi r_{H}^{2}}\left( \sqrt{%
\frac{\Theta }{\pi }}+\frac{4\Theta }{\pi r_{H}}\right) \left( 1-\frac{%
\sigma }{r_{H}^{3\omega _{q}+1}}\right) .  \label{t1}
\end{equation}%
When $\Theta =0$ , Eq. (\ref{t1}) simplifies to the same form found in \cite%
{BB1}. 
\begin{equation}
T_{H}(\omega _{q})=\frac{1}{{4}\pi r_{H}}\left( 1+\frac{3\sigma \omega _{q}}{%
r_{H}^{3\omega _{q}+1}}\right) ,
\end{equation}%
Furthermore, in the absence of the quintessence matter the noncommutative
corrected Hawking temperature reduces to 
\begin{equation}
T_{H}(\Theta )=\frac{1}{{4}\pi r_{H}}-\frac{{1}}{\pi r_{H}^{2}}\left( \sqrt{%
\frac{\Theta }{\pi }}+\frac{4\Theta }{\pi r_{H}}\right) .
\end{equation}%
Finally, for the set of $\Theta =\sigma =0$, the conventional result appears.

\begin{figure}[tbh]
\begin{minipage}[t]{0.5\textwidth}
        \centering
        \includegraphics[width=\textwidth]{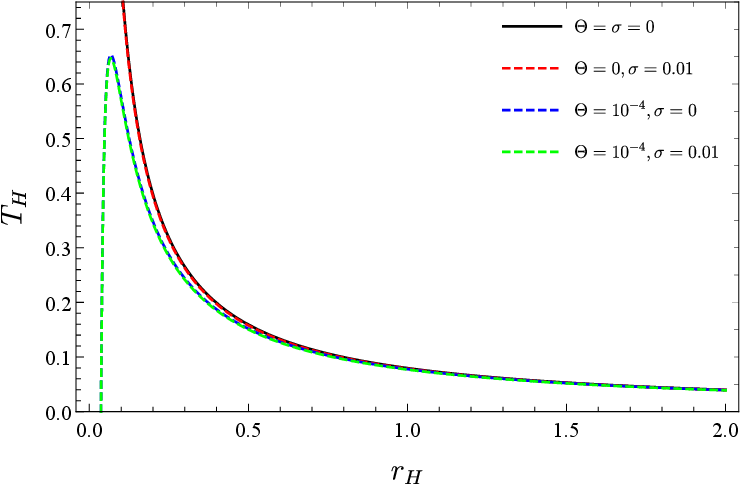}
       \subcaption{ $\omega_{q}=-0.35$.}\label{fig:tta}
   \end{minipage}%
\begin{minipage}[t]{0.5\textwidth}
        \centering
        \includegraphics[width=\textwidth]{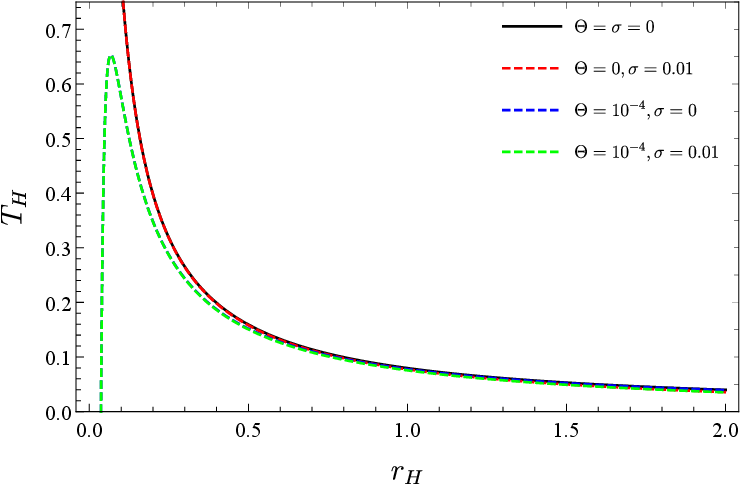}
         \subcaption{  $\omega_{q}=-0.95$.}\label{fig:ttb}
   \end{minipage}
\caption{Hawking temperature versus event horizon radius of the
noncommutative black hole surrounded by quintessence matter.}
\label{fig93}
\end{figure}

To get an appropriate description of the Hawking temperature behavior, in
Figure \ref{fig93} we illustrate the relationship between $T_{H}$ and $r_{H}$
for $\omega _{q}=-0.35$ and $\omega _{q}=-0.95$, respectively. 

\newpage
{ In ordinary case, as $r_{H}\rightarrow 0$, the Hawking temperature becomes infinite. This divergence of the Hawking temperature is known as the \emph{divergence problem} or \emph{infinite temperature problem}. This divergence implies that the black hole would emit an infinite amount of radiation as the black hole approaches to the Planck length, and the semiclassical approximation used to derive the Hawking radiation breaks down. It suggests that a full theory of quantum gravity, which should describe such extreme regimes, is needed to understand the behavior of black holes at these scales. Here, we see that the noncommutativity eliminates this divergence problem of the Hawking temperature. We note that the black hole temperature rises during its evaporation, and it reaches a peak value $T_{H}^{\max }$ at a critical horizon radius value, and subsequently it decreases to zero rapidly. }

\begin{figure}[htb]
\begin{minipage}[t]{0.5\textwidth}
        \centering
        \includegraphics[width=\textwidth]{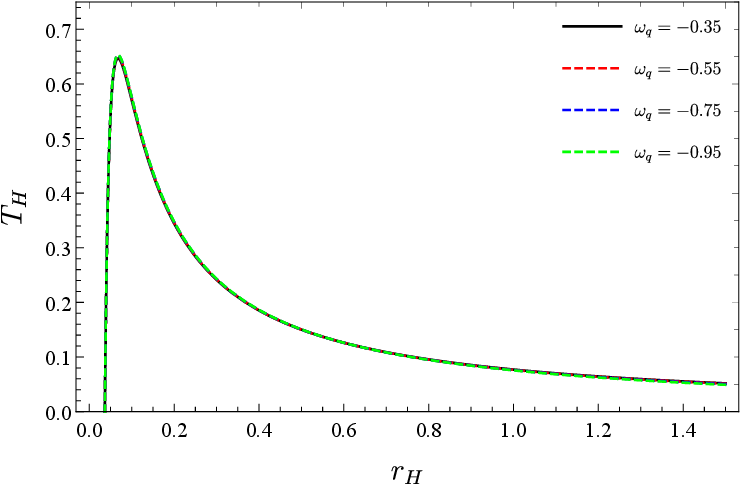}
       \subcaption{ $\sigma=0.01$, and $\Theta=10^{-4}$.}\label{fig:ta}
   \end{minipage}%
\begin{minipage}[t]{0.5\textwidth}
        \centering
        \includegraphics[width=\textwidth]{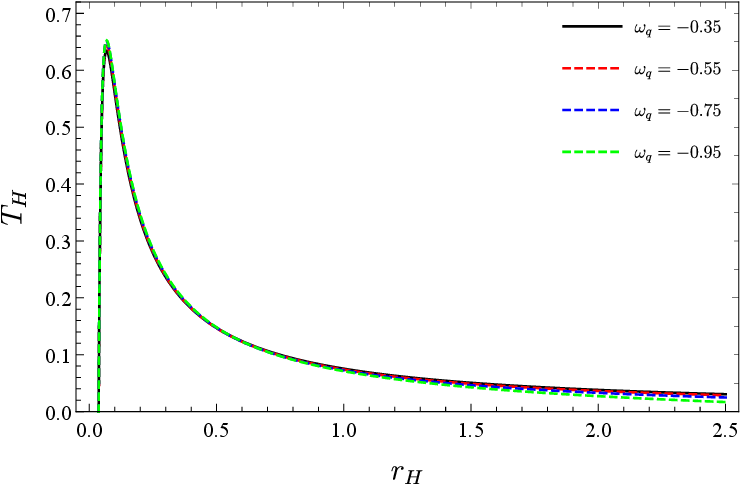}
         \subcaption{ $\sigma=0.03$, and $\Theta=10^{-4}$.}\label{fig:tb}
   \end{minipage}\ 
\begin{minipage}[t]{0.5\textwidth}
        \centering
        \includegraphics[width=\textwidth]{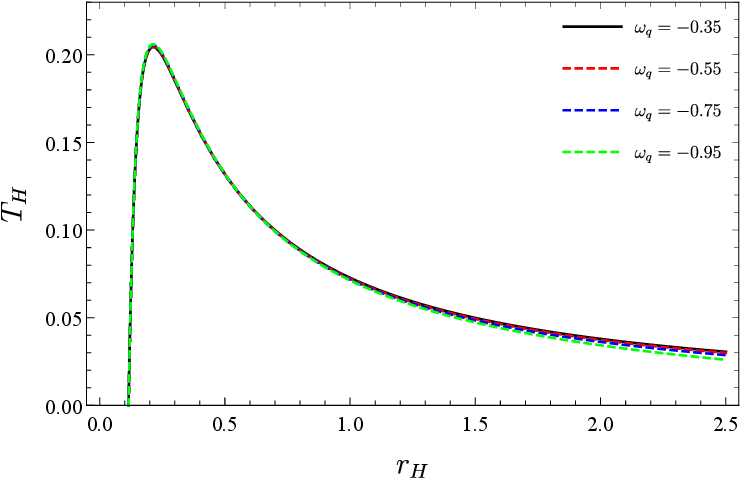}
       \subcaption{ $\sigma=0.01$, and $\Theta=10^{-3}$.}\label{fig:tc}
   \end{minipage}%
\begin{minipage}[t]{0.5\textwidth}
        \centering
        \includegraphics[width=\textwidth]{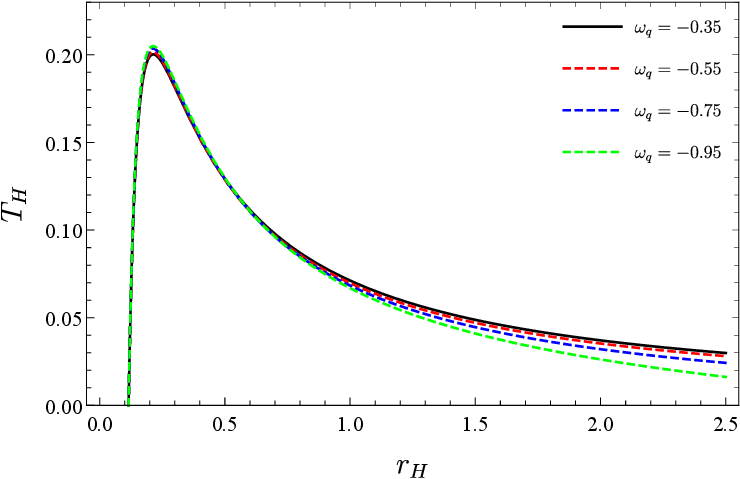}
         \subcaption{ $ \sigma=0.03$, and $\Theta=10^{-3}$.}\label{fig:td}
   \end{minipage}
\caption{The qualitative effects of the noncommutative parameter and
quintessence field on the black hole temperature. }
\label{figtem}
\end{figure}

Then, in Figure \ref{figtem}, we show the impact of the quintessence state
parameter.

\newpage 

We observe that the temperature has a maximum for all the values of $\Theta $
and $\sigma $ such that for fixed $\omega _{q}$ when we increase the value
of $\Theta $ or $\sigma $ the local peak decreases. Next, we utilize the
first law of black hole thermodynamics to derive the Bekenstein entropy,
given by, $dS=\frac{dM}{T}$ \cite{Bardeen, Hawking}. By using Eqs. (\ref{m1}%
) and (\ref{t1}), we get 
{
\begin{eqnarray}
S\left( \Theta \right) \simeq \frac{A}{4}+4\sqrt{\Theta A}+16\Theta\log {\frac{A}{\ell _{p}^{2}}},  \label{s1}
\end{eqnarray}}%
where $A_{H}=4\pi r_{H}^{2}$ is the area of the event horizon, {and $\ell _{p}$ is Planck length}.



\begin{figure}[tbh]
\centering\includegraphics[scale=1]{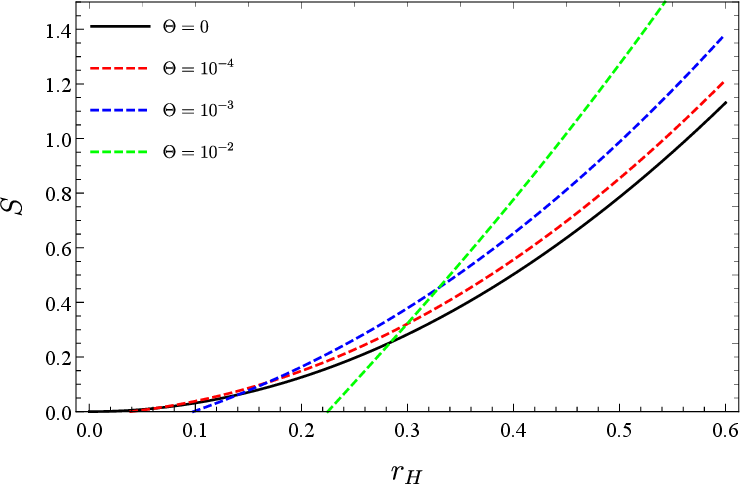}
\caption{The effect of the noncommutative parameter on the entropy function.}
\label{figent}
\end{figure}

\begin{figure}[tbh]
\begin{minipage}[t]{0.5\textwidth}
        \centering
        \includegraphics[width=\textwidth]{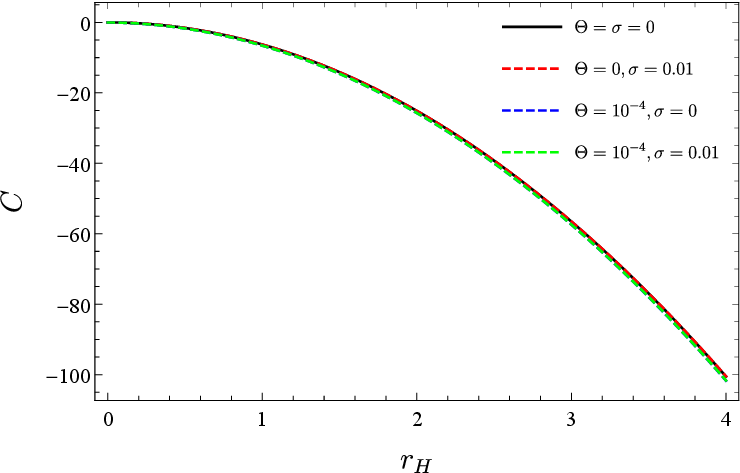}
       \subcaption{ $ \omega_{q}=-0.35$.}\label{fig:c1a}
   \end{minipage}%
\begin{minipage}[t]{0.5\textwidth}
        \centering
        \includegraphics[width=\textwidth]{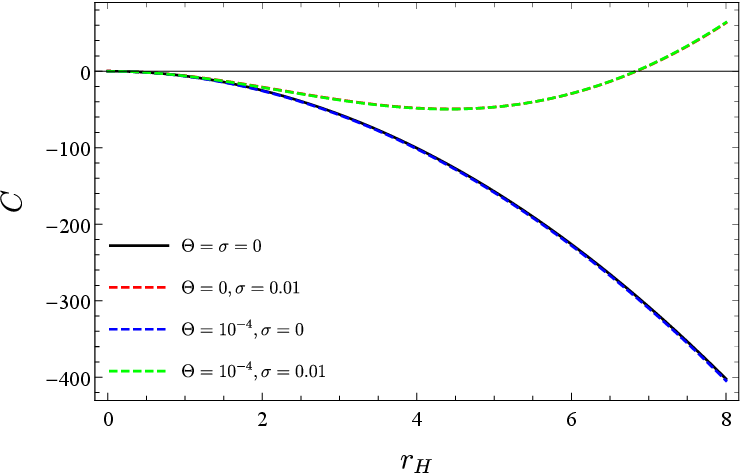}
         \subcaption{ $ \omega_{q}=-0.95$.}\label{fig:c1b}
   \end{minipage}\ 
\caption{Heat capacity function versus event horizon radius of the
noncommutative black hole surrounded by quintessence matter.}
\label{fig39}
\end{figure}

We notice that the noncommutative geometry always modifies the black hole entropy with a positively valued term. Unlike this effect, the presence of
the quintessence matter does not alter the entropy. In Figure \ref{figent},
we plot the noncommutative corrected entropy. 
\newpage We observe that the latter entropy function always takes greater
values than its semi-classical form. { Moreover, we see that depending to the noncommutativity parameter, the entropy become physically meaningful at different horizon.} Next, we employ the following formula
to derive the heat capacity function 
\begin{equation}
C=\frac{dM_{H}}{dT_{H}}.
\end{equation}%
Using Eqs. (\ref{m1}) and (\ref{t1}), we obtain the noncommutative modified
heat capacity in the form of 
\begin{eqnarray}
C(\omega _{q},\Theta ) &=&-\frac{2\pi r_{H}^{2}\left( 1+\frac{3\sigma \omega
_{q}}{r_{H}^{3\omega _{q}+1}}\right) }{1+\frac{3\sigma \omega _{q}(3\omega
_{q}+2)}{r_{H}^{3\omega _{q}+1}}}-\frac{16\sqrt{\pi \Theta }r_{H}\left( 1+%
\frac{\sigma (3\omega _{q}-1)}{r_{H}^{3\omega _{q}+1}}+\frac{3\sigma
^{2}\omega _{q}\left( 9\omega _{q}^{2}+6\omega _{q}-1\right) }{%
2r_{H}^{6\omega _{q}+2}}\right) }{\left( 1+\frac{3\sigma \omega _{q}(3\omega
_{q}+2)}{r_{H}^{3\omega +1}}\right) ^{2}}  \notag \\
&&-\frac{192\Theta \left( 1+\frac{\sigma \left( 3\omega _{q}^{2}+7\omega
_{q}-4\right) }{2r_{H}^{3\omega _{q}+1}}-\frac{\sigma ^{2}\left( 27\omega
_{q}^{4}-18\omega _{q}^{3}-15\omega _{q}^{2}+10\omega _{q}-2\right) }{%
2r_{H}^{6\omega _{q}+2}}+\frac{3\sigma ^{3}\omega _{q}\left( 27\omega
_{q}^{4}+36\omega _{q}^{3}+3\omega _{q}^{2}-5\omega _{q}+1\right) }{%
2r_{H}^{9\omega _{q}+3}}\right) }{\left( 1+\frac{3\sigma \omega _{q}(3\omega
_{q}+2)}{r_{H}^{3\omega _{q}+1}}\right) ^{3}}.\,\,\,\,\,\,
\end{eqnarray}%
{ For $\Theta=\sigma=0$ the heat capacity function reduces to the well-know result,  $C=-2\pi r_{H}^{2}$ \cite{Toghrai2023}.} We see that the heat capacity of the black hole is corrected by both
parameters in a very complex structure. Therefore, we have to use numerical
methods to analyze the effects of additive terms. To this end, we plot
Figure \ref{fig39} and illustrate the heat capacity's behavior for two
different values of the quintessence state parameter. These figures show us
the quintessence state parameter value has a critical role in black hole
stability. For example, for the case $\omega _{q}=-0.35$ the BH is unstable,
however, for $\omega _{q}=-0.95$ the heat capacity can be equal to zero at $%
r_{H}=r_{rem}$ so that black hole terminates the radiation and a remnant
mass occurs. Moreover, for $r_{H}<r_{rem}$ the black hole has a negative
heat capacity which signifies an unstable phase of the black hole.
Similarly, for $r_{H}>r_{rem}$, the black hole becomes stable since it has a
positive heat capacity. 

\begin{figure}[htb]
\begin{minipage}[t]{0.5\textwidth}
        \centering
        \includegraphics[width=\textwidth]{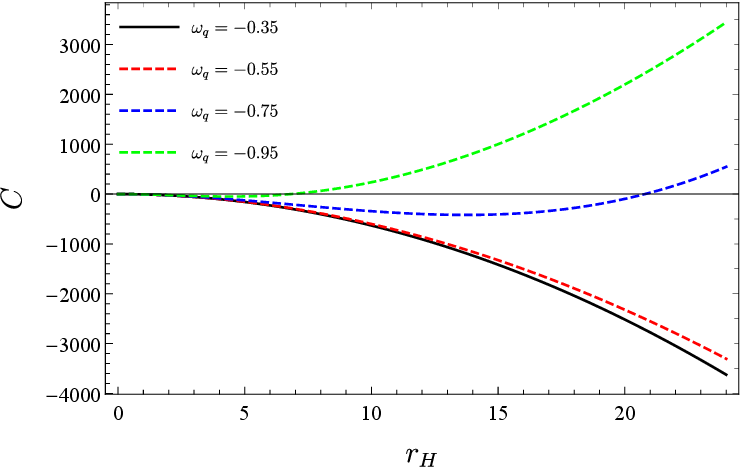}
       \subcaption{ $ \sigma=0.01$, and $\Theta=10^{-4}$.}\label{fig:ca}
   \end{minipage}%
\begin{minipage}[t]{0.5\textwidth}
        \centering
        \includegraphics[width=\textwidth]{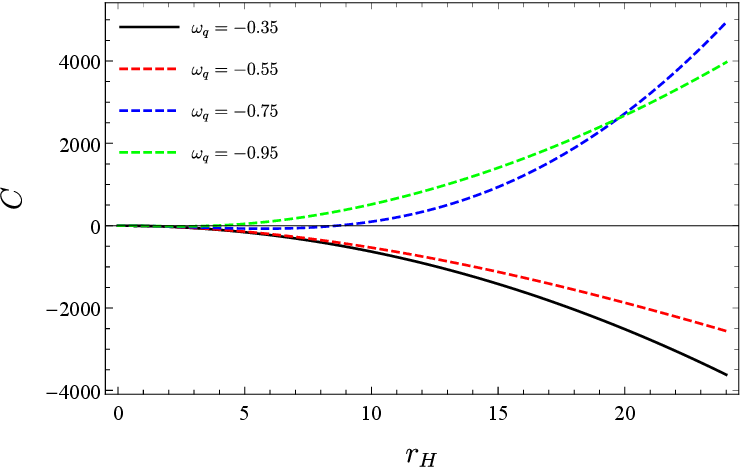}
         \subcaption{ $ \sigma=0.03$, and $\Theta=10^{-4}$.}\label{fig:cb}
   \end{minipage}\ 
\begin{minipage}[t]{0.5\textwidth}
        \centering
        \includegraphics[width=\textwidth]{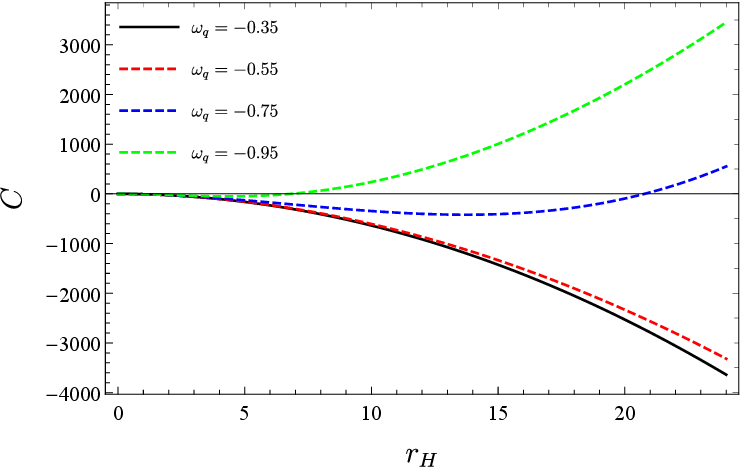}
         \subcaption{ $ \sigma=0.01$, and $\Theta=10^{-3}$.}\label{fig:cc}
   \end{minipage}%
\begin{minipage}[t]{0.5\textwidth}
        \centering
        \includegraphics[width=\textwidth]{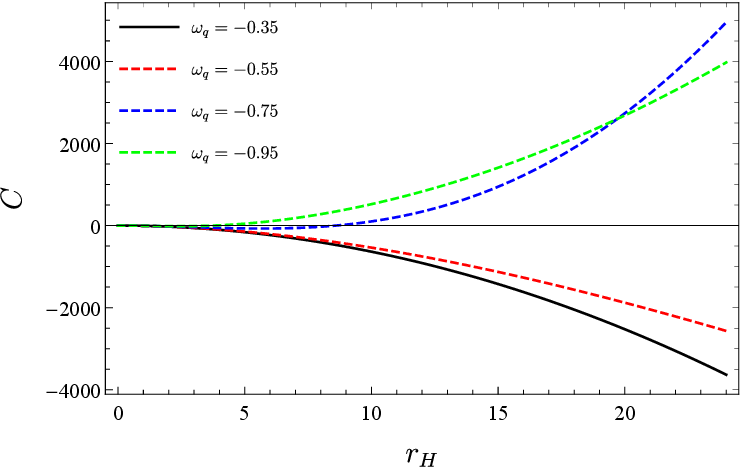}
         \subcaption{ $ \sigma=0.03$, and $\Theta=10^{-3}$.}\label{fig:cd}
   \end{minipage}
\caption{The influence of the noncommutative and quintessence state
parameters on the heat capacity.}
\label{figheat}
\end{figure}

\newpage Figure \ref{figheat} presents the impact of the quintessence field
and noncommutative parameters on the black heat capacity. Similar to the
mass and Hawking temperature cases, we use a set of two different valued
noncommutative and normalization constant parameters. Moreover, in each
plot, we employ the same four different quintessence state parameters which
correspond to four distinct scenarios.

\begin{figure}[tbh]
\begin{minipage}[t]{0.5\textwidth}
        \centering
        \includegraphics[width=\textwidth]{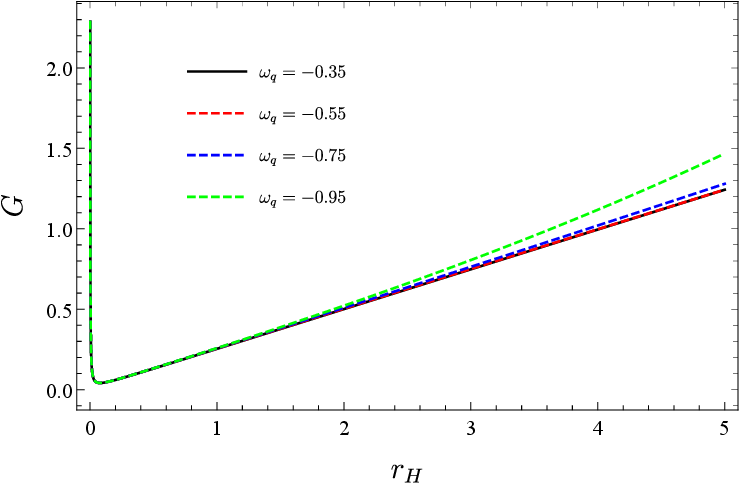}
       \subcaption{ $ \sigma=0.01$, and $\Theta=10^{-4}$.}\label{fig:pa}
   \end{minipage}%
\begin{minipage}[t]{0.5\textwidth}
        \centering
        \includegraphics[width=\textwidth]{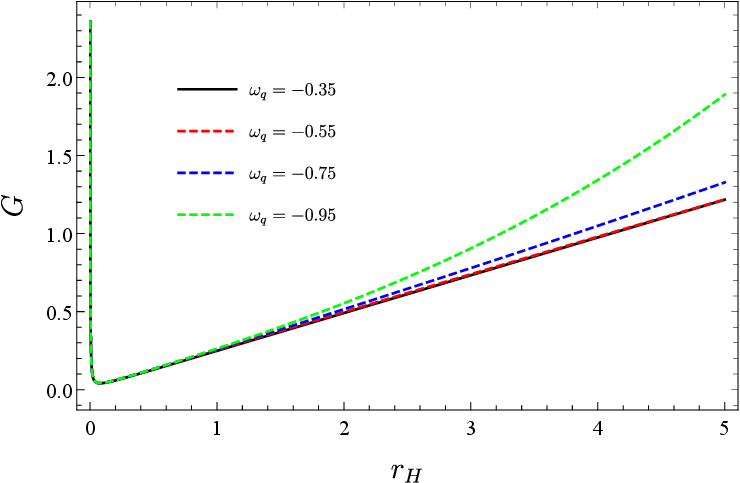}
         \subcaption{ $ \sigma=0.03$, and $\Theta=10^{-4}$.}\label{fig:pb}
   \end{minipage}\ 
\begin{minipage}[t]{0.5\textwidth}
        \centering
        \includegraphics[width=\textwidth]{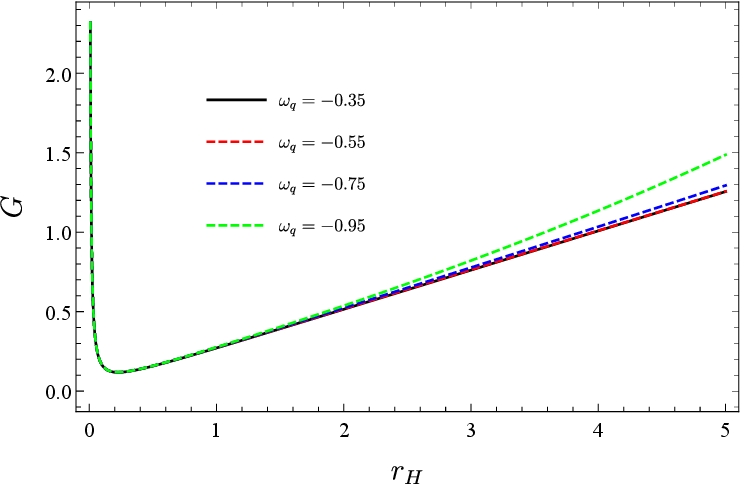}
         \subcaption{ $ \sigma=0.01$, and $\Theta=10^{-3}$.}\label{fig:pc}
   \end{minipage}%
\begin{minipage}[t]{0.5\textwidth}
        \centering
        \includegraphics[width=\textwidth]{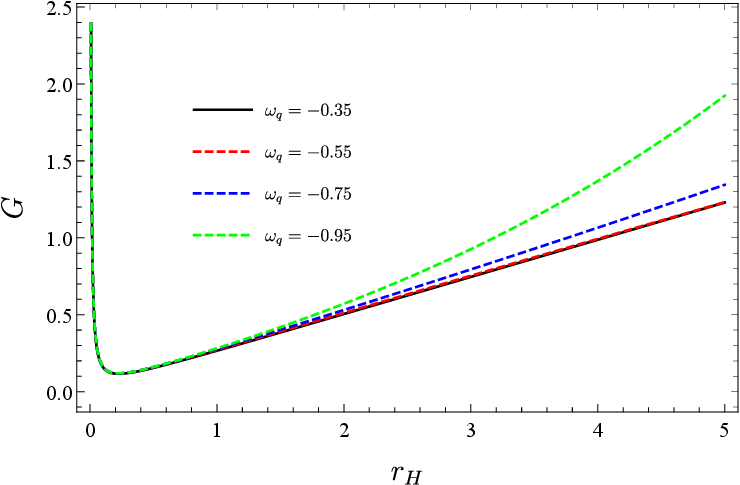}
         \subcaption{ $ \sigma=0.03$, and $\Theta=10^{-3}$.}\label{fig:pd}
   \end{minipage}
\caption{Gibbs free energy function versus event horizon radius of the
noncommutative black hole surrounded by quintessence matter.}
\label{figgib}
\end{figure}

\newpage We observe that the effect of the noncommutative parameter and
quintessence field are very significant. For additional details on the phase
transition and overall stability, our focus turns to the Gibbs free energy,
which is defined by 
\begin{equation}
G=M-TS,  \label{g1}
\end{equation}%
Substituting Eqs. (\ref{m1}), (\ref{t1}), and (\ref{s1}) into Eq. (\ref{g1}%
), we obtain 
\begin{eqnarray}
G(\omega _{q},\Theta ) &=&\frac{r_{H}}{4}\left( 1-\frac{\sigma (3\omega
_{q}+2)}{r_{H}^{1+3\omega _{q}}}\right) +\sqrt{\frac{\Theta }{\pi }}\left( 1-%
\frac{3\sigma (2\omega _{q}+1)}{r_{H}^{1+3\omega _{q}}}\right)   \notag \\
&&-\frac{4\Theta }{\pi }\left( \frac{2}{r_{H}}\left( 1+\frac{3\sigma \omega
_{q}}{r_{H}^{3\omega _{q}+1}}\right) \log {\frac{r_{H}}{\ell_p}}-\frac{5}{r_{H}}\left( 1-%
\frac{\sigma }{r_{H}^{3\omega _{q}+1}}\right) \right) .
\end{eqnarray}%
In Figure \ref{figgib}, we depict the Gibbs free energy function versus
event horizon. 

\newpage These figures show that the Gibbs free energy function has a
turning point which can be determined via $\frac{\partial G}{\partial r_{H}}%
=0$. Here, we observe that the location of that minimum value depends on the
normalization factor associated with the quintessence field and
noncommutative parameters.


\section{Shadow in the presence of plasma}

\label{sec4}

\begin{figure}[htb]
\begin{minipage}[t]{0.50\textwidth}
        \centering
        \includegraphics[width=\textwidth]{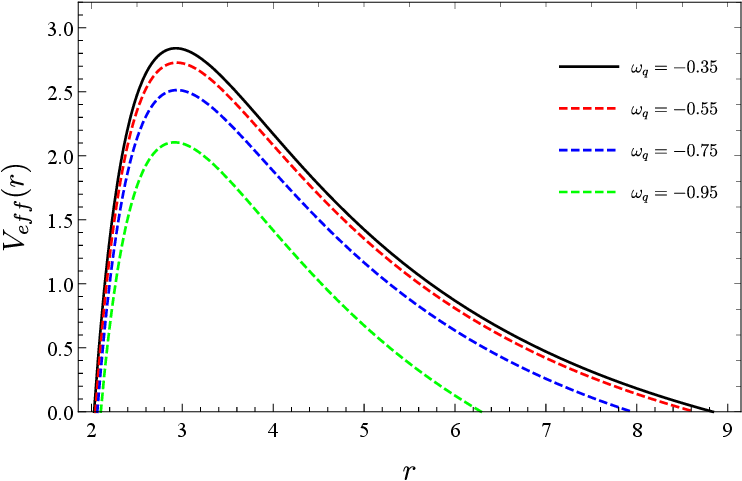}
       \subcaption{ $\Theta=10^{-3}${,} $ \sigma=0.01$.}\label{fig:va}
   \end{minipage}%
\begin{minipage}[t]{0.50\textwidth}
        \centering
        \includegraphics[width=\textwidth]{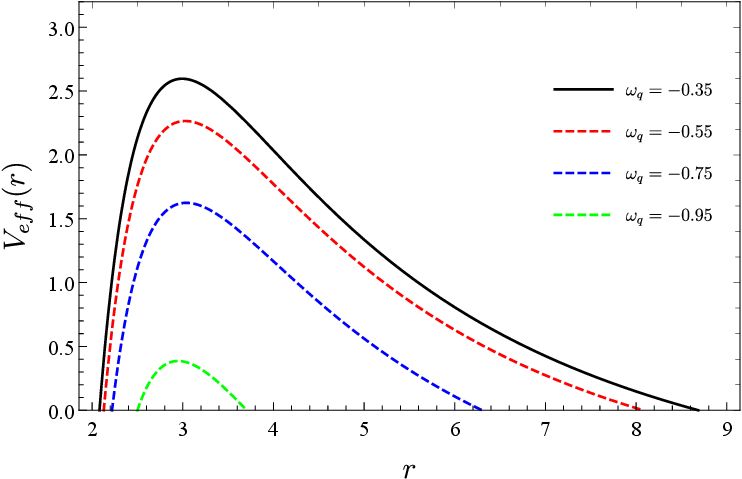}
         \subcaption{ $\Theta=10^{-3}${,} $ \sigma=0.03$.}\label{fig:vb}
   \end{minipage}\\
\begin{minipage}[t]{0.50\textwidth}
       \centering
        \includegraphics[width=\textwidth]{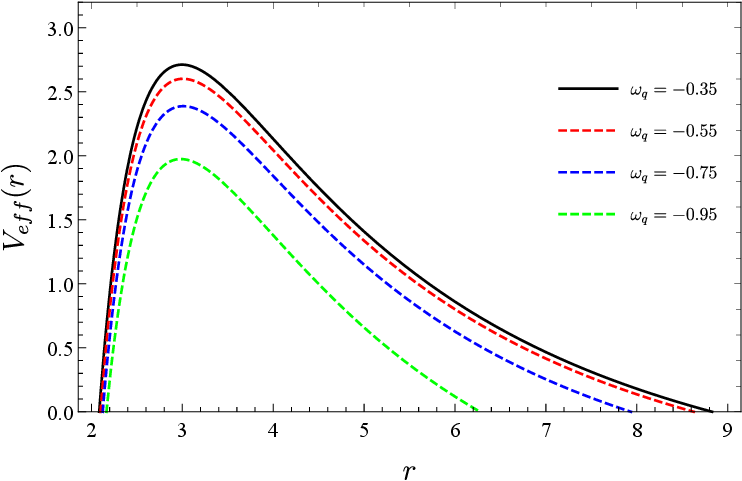}
       \subcaption{$\Theta=10^{-4}${,} $ \sigma=0.01$.}\label{fip:vc}
   \end{minipage}%
\begin{minipage}[t]{0.50\textwidth}
        \centering
        \includegraphics[width=\textwidth]{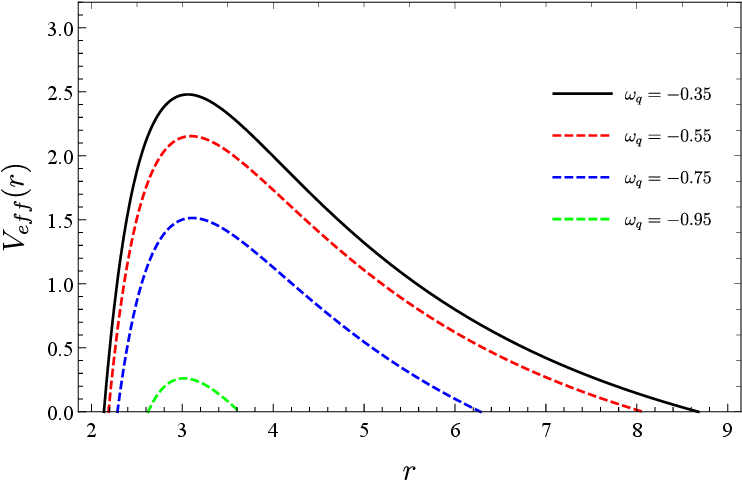}
       \subcaption{$\Theta=10^{-4}${,} $ \sigma=0.03$.}\label{fip:vd}
   \end{minipage}
\caption{The variation of the effective potential as a function of the
radial coordinate for two values of   {$\Theta $ and $\sigma $   with} $L=10$, $\mathcal{K=}1$, $E=1$, $k=0.1$. }
\label{Veff}
\end{figure}

\begin{figure}[htb]
\begin{minipage}[t]{0.50\textwidth}
        \centering
        \includegraphics[width=\textwidth]{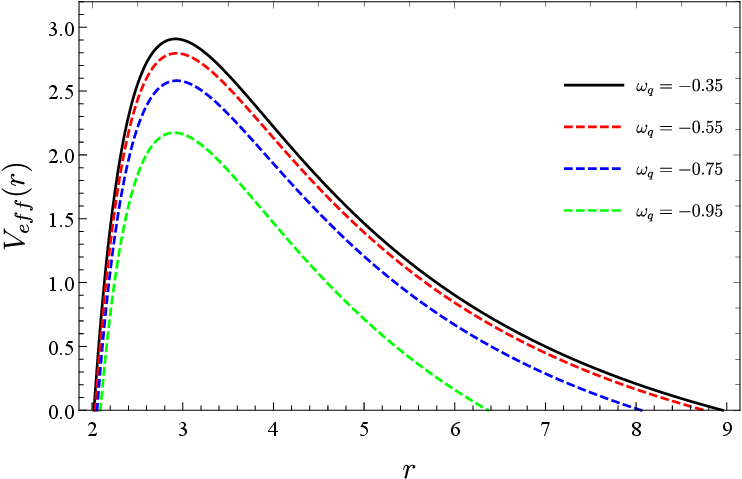}
       \subcaption{ $\Theta=10^{-3}${,} $ \sigma=0.01$.}\label{fig:vaa}
   \end{minipage}%
\begin{minipage}[t]{0.50\textwidth}
        \centering
        \includegraphics[width=\textwidth]{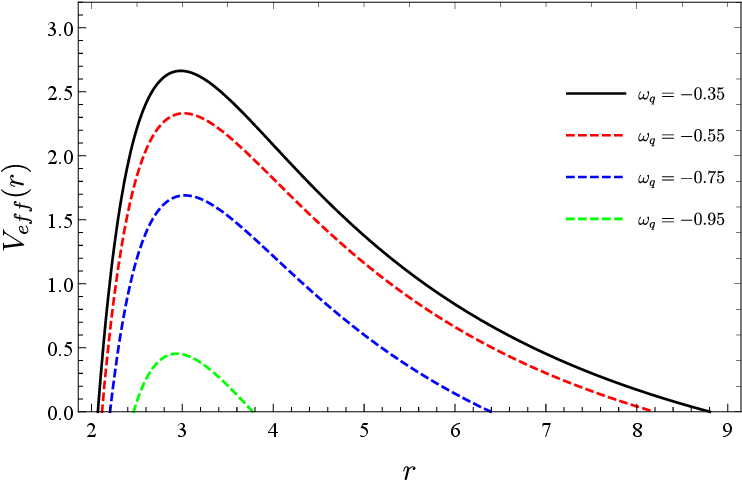}
         \subcaption{ $\Theta=10^{-3}${,} $ \sigma=0.03$.}\label{fig:vvb}
   \end{minipage}\\ 
\begin{minipage}[t]{0.50\textwidth}
       \centering
        \includegraphics[width=\textwidth]{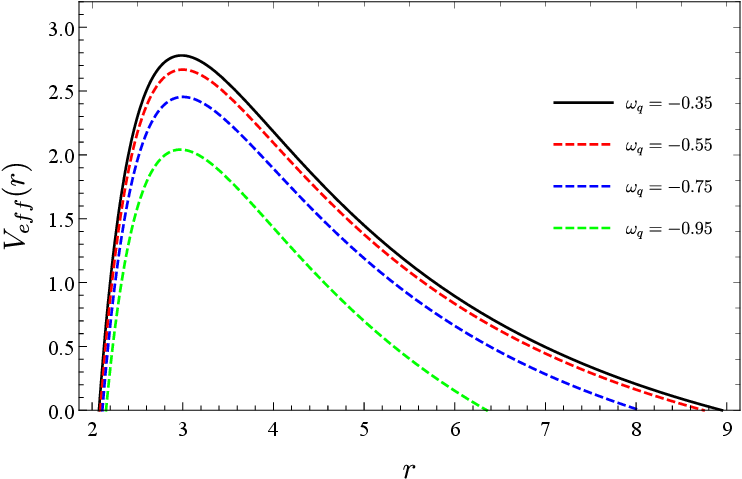}
       \subcaption{$\Theta=10^{-4}${,} $ \sigma=0.01$.}\label{fip:vvc}
   \end{minipage}%
\begin{minipage}[t]{0.50\textwidth}
        \centering
        \includegraphics[width=\textwidth]{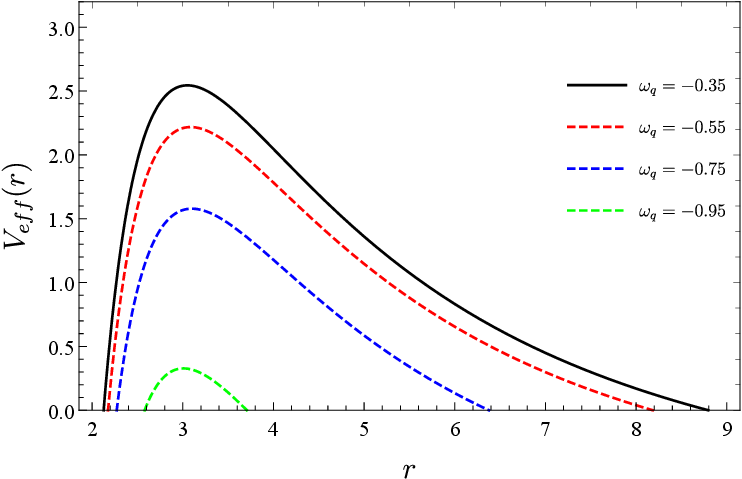}
       \subcaption{$\Theta=10^{-4}${,} $ \sigma=0.03$.}\label{fip:vvd}
   \end{minipage}
\caption{The variation of the effective potential as a function of the
radial coordinate for two values of \ $\Theta $ { and $\sigma$ with} $L=10$, $\mathcal{K=}1$,  $E=1$, $k=0.3$.}
\label{veffk}
\end{figure}

In this section, we assume a plasma distribution that surrounds the black
hole with a refractive index, $n=n\left( x,\omega \right) $. Here, $\omega $
corresponds to the photon frequency measured by an observer that moves with
a velocity $u^{\alpha }$. In this case, the effective energy of the photon
reads: 
\begin{equation}
\hbar \omega =-p_{\mu }u^{\mu }.
\end{equation}
In \cite{Roger}, the refractive index of the medium is given as a function
of the photon four-momentum 
\begin{equation}
n^{2}=1+\frac{p_{\mu }p^{\mu }}{\left( p_{\mu }u^{\mu }\right) ^{2}}.
\end{equation}%
In the vacuum scenario, $n=1$, the standard condition for null geodesics is
restored by $p_{\mu }p^{\mu }=0$. Now, we consider a photon around a given
spacetime geometry surrounding plasma. We express the Hamiltonian, as given
in \cite{Roger} 
\begin{equation}
H=\frac{1}{2}\left[ g^{\mu \nu }p_{\mu }p_{\nu }-\left( n^{2}-1\right)
\left( p_{0}\sqrt{-g^{00}}\right) ^{2}\right] .
\end{equation}
In \cite{Roger, Abdujabbarov}, authors noted that one can introduce a
particular expression for the plasma frequency to facilitate analytical
computations, under the assumption that the refractive index follows a
general form, 
\begin{eqnarray}
n^{2}=1-\frac{\omega _{e}^{2}}{\omega ^{2}}=1-\frac{k}{r},\text{ \ \ \ \ }%
k>0.  \label{n}
\end{eqnarray}
Now, we have to introduce how to obtain the trajectories of photons. In
terms of the affine parameter, $\tau$, they are governed by the following
set of equations: 
\begin{eqnarray}
\frac{\partial x^{\mu }}{\partial \tau }&=&\dot{x}^{\mu }=\frac{\partial }{%
\partial p_{\mu }}H, \\
\frac{\partial p_{\mu }}{\partial \tau }&=&\dot{p}_{\mu }=-\frac{\partial }{%
\partial x^{\mu }}H.
\end{eqnarray}
After simple manipulation, we write the components of the canonically
conjugate momentum as follows: 
\begin{eqnarray}
E&=&p_{0}=-\frac{f\left( r\right) }{n^{2}}\,\, \dot{t},  \label{t} \\
L&=&p_{\phi }=r^{2}\sin ^{2}\theta \,\, \dot{\phi}.  \label{phi}
\end{eqnarray}
Here, $E$ and $L$  represent the { energy and angular momentum of the photons,} respectively. { Then, 
we deduce} the remaining two geodesic equations
from the Hamilton-Jacobi equation, 
\begin{equation}
\frac{\partial }{\partial \tau }S=-\frac{1}{2}\left[ g^{\mu \nu }\frac{%
\partial S}{\partial x^{\mu }}\frac{\partial S}{\partial x^{\nu }}-\left(
n^{2}-1\right) \left( \frac{\partial S}{\partial t}\sqrt{-g^{00}}\right) ^{2}%
\right] .  \label{action}
\end{equation}%
Using Eq. (\ref{metric11}) in Eq. (\ref{action}), we obtain%
\begin{equation}
\frac{\partial }{\partial \tau }S=-\frac{1}{2}\left[ -\frac{n^{2}}{f\left(
r\right) }\left( \frac{\partial S}{\partial t}\right) ^{2}+f\left( r\right)
\left( \frac{\partial S}{\partial r}\right) ^{2}+\frac{1}{r^{2}}\left( \frac{%
\partial S}{\partial \theta }\right) ^{2}+\frac{1}{r^{2}\sin ^{2}\theta }%
\left( \frac{\partial S}{\partial \phi }\right) ^{2}\right] .  \label{act21}
\end{equation}%
{ We then} assume a separable Jacobi action solution 
\begin{equation}
S=-Et+L\phi +S_{\theta }+S_{r},  \label{Jacobi}
\end{equation}%
where $S_{r}$ { component depends only on $r$, while $S_{\theta }$ component relies only on $\theta$.} Then. by replacing Eq. (\ref{Jacobi}) with Eq. (\ref{act21}),
we arrive at 
\begin{equation}
\frac{n^{2}}{f\left( r\right) }E^{2}-f\left( r\right) \left( \frac{\partial
S_{r}}{\partial r}\right) ^{2}-\frac{1}{r^{2}}\left[ \left( \frac{\partial
S_{\theta }}{\partial \theta }\right) ^{2}-\mathcal{K}+L^{2}\cot ^{2}\theta %
\right] -\frac{1}{r^{2}}\left[ \frac{L^{2}}{\sin ^{2}\theta }+\mathcal{K}%
-L^{2}\cot ^{2}\theta \right] =0,  \label{action11}
\end{equation}%
{ where $\mathcal{K}$ is} the Carter constant. { Then, we recast} Eq. (\ref%
{action11}) as the following two separated equations%
\begin{eqnarray}
\left( \frac{\partial S_{\theta }}{\partial \theta }\right) ^{2}&=&\mathcal{K%
}-L^{2}\cot ^{2}\theta ,  \label{26} \\
f^{2}\left( r\right) \left( \frac{\partial S_{r}}{\partial r}\right)
^{2}&=&n^{2}E^{2}-\frac{f\left( r\right) }{r^{2}}\left( L^{2}+\mathcal{K}%
\right) .  \label{27}
\end{eqnarray}
Now, using the relation $\left( \frac{\partial S_{\theta }}{\partial \theta }%
\right) =p_{\theta }$, we obtain 
\begin{equation}
\frac{\partial S_{\theta }}{\partial \theta }=r^{2}\dot{\theta}.  \label{24}
\end{equation}%
Similarly, employing the relation $\left( \frac{\partial S_{r}}{\partial r}
\right) =p_{r}$, we find 
\begin{equation}
\frac{\partial S_{r}}{\partial r}=\frac{\dot{r}}{f\left( r\right) }.
\label{25}
\end{equation}%
Then, we substitute Eqs. (\ref{24}) and (\ref{25}) to Eqs. (\ref{26}) and (%
\ref{27}) to express the complete null geodesic equations 
\begin{eqnarray}
r^{2}\dot{r}&=&\pm \sqrt{n^{2}E^{2}r^{4}-r^{2}f\left( r\right) \left( L^{2}+%
\mathcal{K}\right) }=\pm \sqrt{\mathcal{R}},  \label{radeq11} \\
r^{2}\dot{\theta}&=&\sqrt{\mathcal{K}-L^{2}\cot ^{2}\theta }.  \label{teta11}
\end{eqnarray}
Since the equations of motion rely on conserved quantities $E$, $L$ { and} $\mathcal{K}$, it is advantageous to express them in terms
of normalized parameters 
\begin{eqnarray}
\zeta =\frac{L^{2}}{E^{2}}, \qquad \eta =\frac{\mathcal{K}}{E^{2}}.
\end{eqnarray}
{By doing so}, the radial equation, namely Eq.{,}  (\ref{radeq11}) can be
reformulated into a more common form { of}
\begin{equation}
\left( \frac{\partial r}{\partial \tau }\right) ^{2}+V_{eff}\left( r\right)
=0,
\end{equation}
with the effective radial potential 
\begin{equation}
V_{eff}\left( r\right) =E^{2}\left[ \frac{f\left( r\right) }{r^{2}}\left(
\zeta ^{2}+\eta \right) -n^{2}\right].
\end{equation}%
In Figures (\ref{Veff}) and (\ref{veffk}), we depict the impact of the
photon's radial motion on the effective potential. 

\newpage \noindent Now, we explore the circular photon orbits. By
definitions, they exist with the following two conditions 
\begin{eqnarray}
\left. V_{eff}\left( r\right) \right\vert _{r=r_{p}}&=&0,  \label{peff} \\
\left. \frac{\partial V_{eff}\left( r\right) }{\partial r}\right\vert
_{r=r_{p}}&=&0.
\end{eqnarray}%
Here, we also have to take { the maximizing condition of the effective potential
into account}
\begin{equation}
\left. \frac{\partial ^{2}V_{eff}\left( r\right) }{\partial r^{2}}%
\right\vert _{r=r_{p}}<0.
\end{equation}%
With the help of the first condition, { given} by Eq. (\ref{peff}), we realize that the impact parameters, $\zeta $ and $\eta $, should satisfy the following criteria 
\begin{equation}
\zeta ^{2}+\eta =\frac{r_{p}^{2}\left( 1-\frac{k}{r_{p}}\right) }{1-\frac{2}{%
r_{p}}-\frac{\sigma }{r_{p}^{3\omega _{q}+1}}}-\frac{8\sqrt{\Theta }\left( 1-%
\frac{k}{r_{p}}\right) }{\sqrt{\pi }\left( 1-\frac{2}{r_{p}}-\frac{\sigma }{%
r_{p}^{3\omega _{q}+1}}\right) {}^{2}}+\frac{64\Theta \left( 1-\frac{k}{r_{p}%
}\right) }{\pi r_{p}^{2}\left( 1-\frac{2}{r_{p}}-\frac{\sigma }{%
r_{p}^{3\omega _{q}+1}}\right) {}^{2}},  \label{para}
\end{equation}%
{ while from} the second condition, we { find out} 
\begin{equation}
\Big( n\left( r\right) rf^{\prime }\left( r\right) -2n\left( r\right)
f\left( r\right) -2rn^{\prime }\left( r\right) f\left( r\right) \bigg\vert %
_{r=r_{p}}=0.  \label{photo}
\end{equation}%
{ Here, the prime sign denotes derivative  with respect to $r_{p}$. Then, we utilize $f\left( r\right) $ and $n\left(r\right) $ expressions, given in Eqs. (\ref{metric11}) and (\ref{n}), to} obtain an equation for the radius of the photon sphere. However, the resulting expression is intricate,
and for brevity, we refrain from presenting it here. Therefore, we opt for a
numerical solution. It is worth noting that the inclusion of the plasma
medium introduces an additional parameter, $k$, in Eq. (\ref{photo}). For
the numerical calculations, we consider different values for $\sigma $, $%
\omega _{q}$, $\Theta$, and $k$. Subsequently, we determine the values of
the photon sphere radius and impact parameters numerically by solving Eq.(%
\ref{photo}). We tabulate our results in Tables \ref{tab1} and \ref{tab2}. 
\begin{table}[tbh]
\begin{tabular}{|l|ll|ll|ll|ll|}
\hline
\rowcolor{lightgray} \multirow{2}{*}{} & \multicolumn{2}{|c|}{$\Theta
=10^{-3},\sigma =0.01$} & \multicolumn{2}{|c}{$\Theta =10^{-4},\sigma =0.01$}
& \multicolumn{2}{|c|}{$\Theta =10^{-3},\sigma =0.03$} & 
\multicolumn{2}{|c|}{$\Theta =10^{-4},\sigma =0.03$} \\ \hline\hline
\rowcolor{lightgray} $\omega _{q}$ & $r_{p}$ & $\eta +\zeta ^{2}$ & $r_{p}$
& $\eta +\zeta ^{2}$ & $r_{p}$ & $\eta +\zeta ^{2}$ & $r_{p}$ & $\eta +\zeta
^{2}$ \\ \hline
$-0.35$ & 2.91550 & 25.6359 & 2.98370 & 26.5396 & 2.98006 & 27.4024 & 3.04827
& 28.3481 \\ 
$-0.55$ & 2.92592 & 26.4186 & 2.99471 & 27.3648 & 3.01541 & 30.2196 & 3.08550
& 31.3169 \\ 
$-0.75$ & 2.92935 & 28.0477 & 2.99883 & 29.1102 & 3.02986 & 37.7032 & 3.10239
& 39.3848 \\ 
$-0.95$ & 2.90408 & 31.7756 & 2.97306 & 33.1934 & 2.94601 & 72.5113 & 3.01670
& 79.5340 \\ \hline\hline
\end{tabular}
\label{tabm}
\caption{{}Numerical estimations of the photon radius and the impact
parameters for $k=0.1.$}
\label{tab1}
\end{table}

\begin{table}[tbh]
\begin{tabular}{|l|ll|ll|ll|ll|}
\hline
\rowcolor{lightgray} \multirow{2}{*}{} & \multicolumn{2}{|c|}{$\Theta
=10^{-3},\sigma =0.01$} & \multicolumn{2}{|c}{$\Theta =10^{-4},\sigma =0.01$}
& \multicolumn{2}{|c|}{$\Theta =10^{-3},\sigma =0.03$} & 
\multicolumn{2}{|c|}{$\Theta =10^{-4},\sigma =0.03$} \\ \hline\hline
\rowcolor{lightgray} $\omega _{q}$ & $r_{p}$ & $\eta +\zeta ^{2}$ & $r_{p}$
& $\eta +\zeta ^{2}$ & $r_{p}$ & $\eta +\zeta ^{2}$ & $r_{p}$ & $\eta +\zeta
^{2}$ \\ \hline
$-0.35$ & 2.87819 & 23.8033 & 2.94693 & 24.6870 & 2.94287 & 25.4877 & 3.01160
& 26.4130 \\ 
$-0.55$ & 2.88902 & 24.5373 & 2.95834 & 25.4621 & 2.97953 & 28.1344 & 3.05014
& 29.2064 \\ 
$-0.75$ & 2.89410 & 26.0535 & 2.96413 & 27.0896 & 2.99960 & 35.1177 & 3.07271
& 36.7481 \\ 
$-0.95$ & 2.87385 & 29.4984 & 2.94350 & 30.8707 & 2.93166 & 67.4244 & 3.00336
& 74.0682 \\ \hline\hline
\end{tabular}
\label{tabm2}
\caption{{}Numerical estimations of the photon radius and the impact
parameters for $k=0.3.$}
\label{tab2}
\end{table}

\begin{figure}[!htb]
\begin{minipage}[t]{0.50\textwidth}
        \centering
        \includegraphics[width=\textwidth]{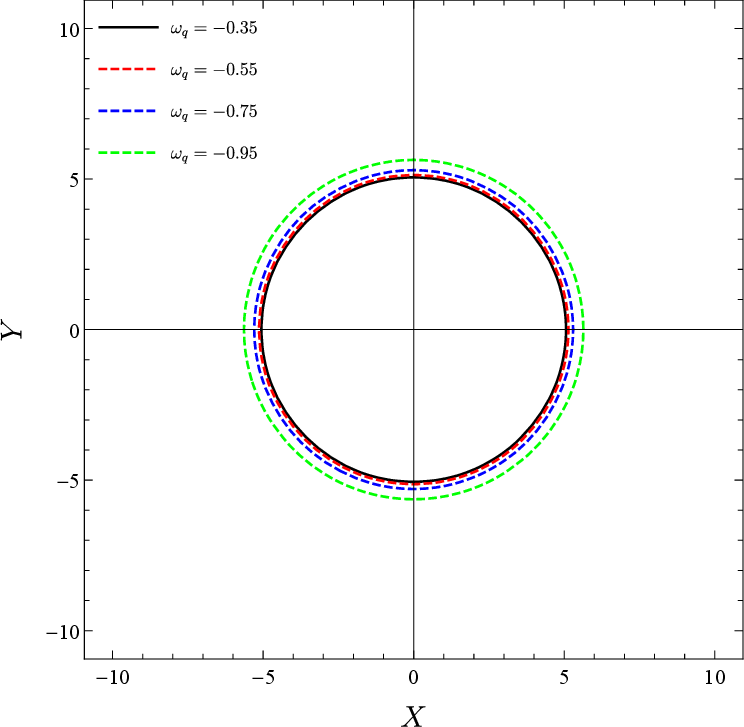}
       \subcaption{ $ \sigma=0.01$ and $\Theta=10^{-3}$.}\label{fig:pla}
   \end{minipage}%
\begin{minipage}[t]{0.50\textwidth}
       \centering
        \includegraphics[width=\textwidth]{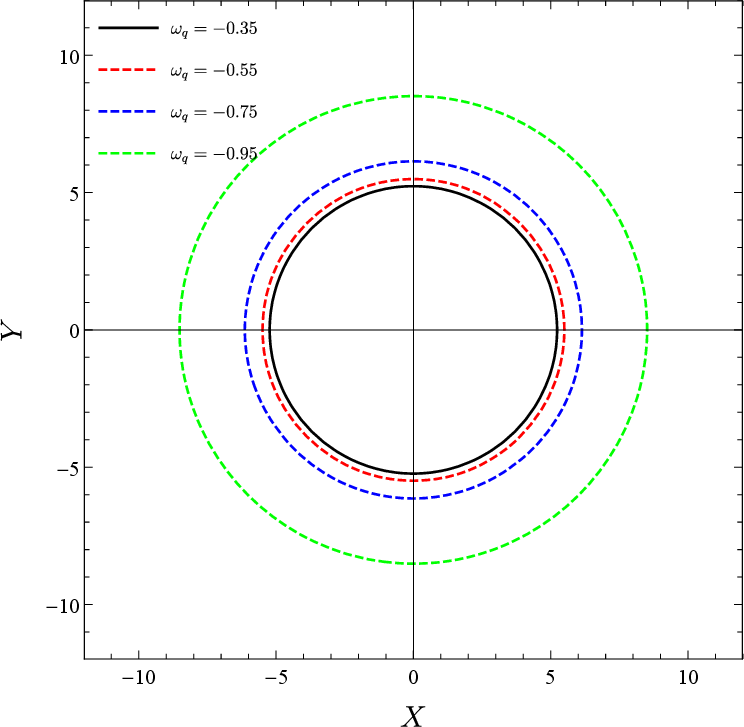}
       \subcaption{ $ \sigma=0.03$ and $\Theta=10^{-3}$.}\label{fig:plb}
   \end{minipage}\\
   \begin{minipage}[t]{0.50\textwidth}
        \centering
        \includegraphics[width=\textwidth]{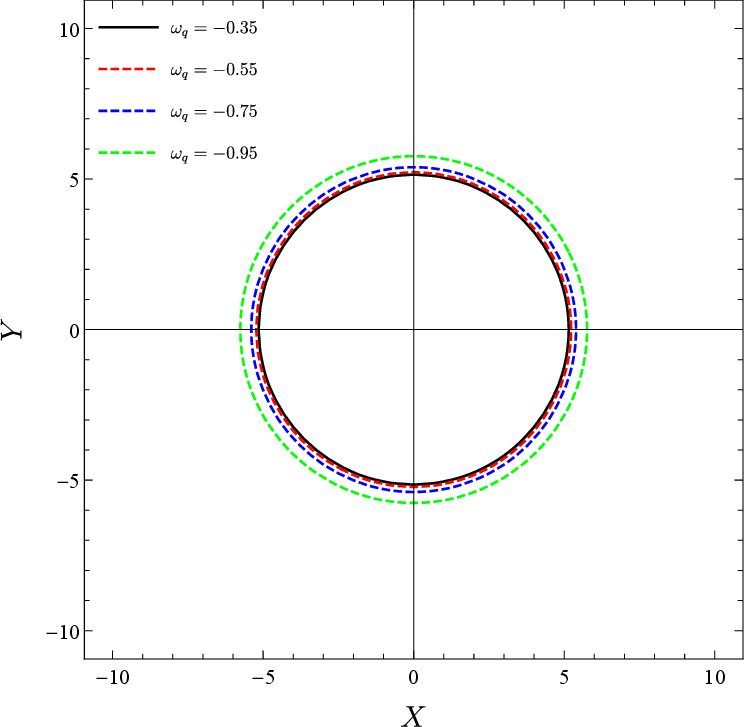}
         \subcaption{ $ \sigma=0.01$ and $\Theta=10^{-4}$.}\label{fig:plc}
   \end{minipage}
\begin{minipage}[t]{0.50\textwidth}
        \centering
        \includegraphics[width=\textwidth]{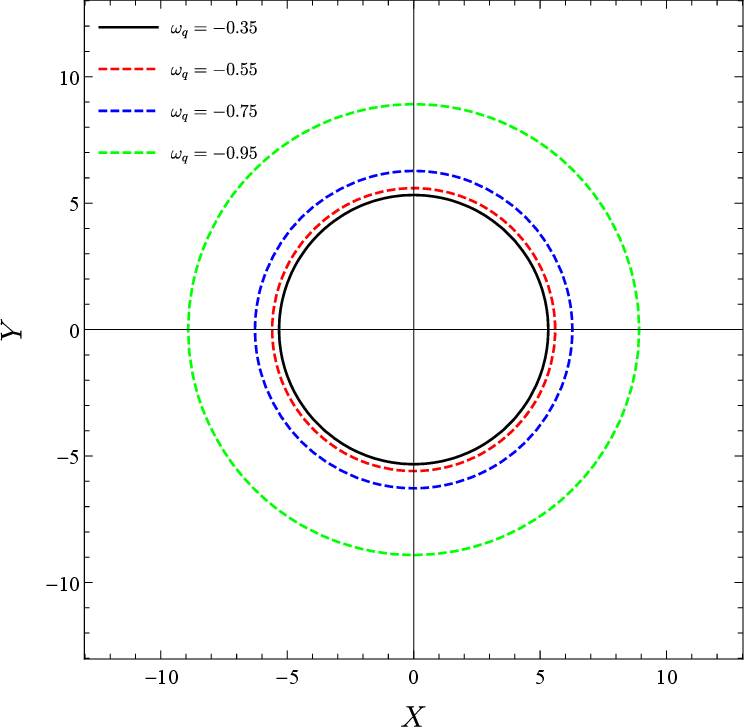}
       \subcaption{ $ \sigma=0.03$ and $\Theta=10^{-4}$.}\label{fip:pld}
   \end{minipage}
\caption{Black hole shadow in the celestial plane ($X$-$Y$) for different values of $\omega _{q}$ and $k=0.1$.}
\label{figsh1}
\end{figure}

{ We are now focusing on obtaining the black hole shadow with geodesic equations and criteria for unstable circular orbits.} To facilitate this examination, we employ the celestial coordinates ($X$, $Y$)
defined as \cite{Papnoi}: 
\begin{eqnarray}
X&=&\lim_{r_{o}\rightarrow \infty }\left( -r_{o}^{2}\sin \theta _{o}\frac{%
d\phi }{dr}\right) , \\
Y&=&\lim_{r_{o}\rightarrow \infty }\left( r_{o}\frac{d\theta }{dr}\right) ,
\end{eqnarray}
{where} $r_{o}$ corresponds to the distance between the observer and the black hole, {while} $\theta _{o}$ {represents} the angle of inclination between the observer's line of sight and the rotation axis of the black hole. Here, the values of $\frac{d\phi }{dr}$ and $\frac{d\theta }{dr}$\ can be determined
by the geodesic equations given in Eqs. (\ref{phi}), (\ref{radeq11}) and (%
\ref{teta11}): 
\begin{eqnarray}
\frac{d\phi }{dr}&=&\frac{\zeta }{r^{2}\sin ^{2}\theta \sqrt{n^{2}-\frac{%
f\left( r\right) }{r^{2}}\left( \zeta ^{2}+\eta \right) }}, \\
\frac{d\theta }{dr}&=&\frac{1}{r^{2}}\frac{\sqrt{\eta -\zeta ^{2}\cot
^{2}\theta }}{\sqrt{n^{2}-\frac{f\left( r_{o}\right) }{r^{2}}\left( \zeta
^{2}+\eta \right) }}.
\end{eqnarray}%
Substituting these equations into the definitions of $X$ and $Y$, and taking
the limit $r_{o}\rightarrow \infty $, we acquire: 
\begin{eqnarray}
X &=& -\zeta , \\
Y &=& \sqrt{\eta -\zeta ^{2}\cot ^{2}\theta _{o}} .
\end{eqnarray}
Next, for simplicity, we assume the observer on the equatorial plane, thus $%
\theta _{o}=\frac{\pi }{2}$. Then, these equations undergo simplification. 
\begin{equation}
X=-\zeta, \qquad Y=\sqrt{\eta }.  \label{obs}
\end{equation}%
Using Eq. (\ref{para}), one can read Eq. (\ref{obs}) as%
\begin{equation}
X^{2}+Y^{2}=\zeta ^{2}+\eta =R_{s}^{2}.  \label{Rs}
\end{equation}%
Here, the quantity $R_{s}$ is known as the radius of the shadow. { Now, we aim to demonstrate the impact of the fuzziness and the quintessence matter parameters on the shadow of the black hole. First, we depict Figure \ref{figsh1}. }  

\begin{figure}[!htb]
\begin{minipage}[t]{0.50\textwidth}
        \centering
        \includegraphics[width=\textwidth]{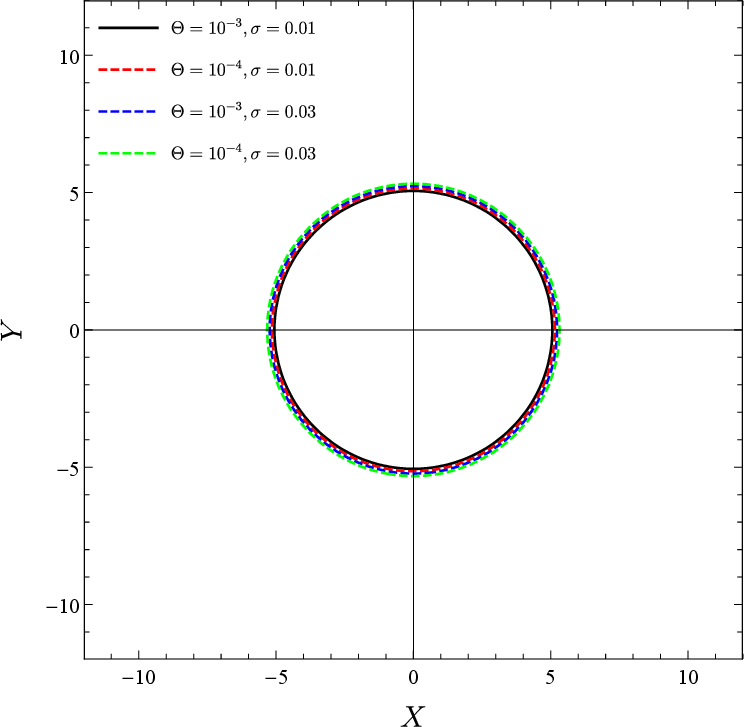}
       \subcaption{ $ \omega_{q}=-0.35$.}\label{fig:sha}
   \end{minipage}%
\begin{minipage}[t]{0.50\textwidth}
        \centering
        \includegraphics[width=\textwidth]{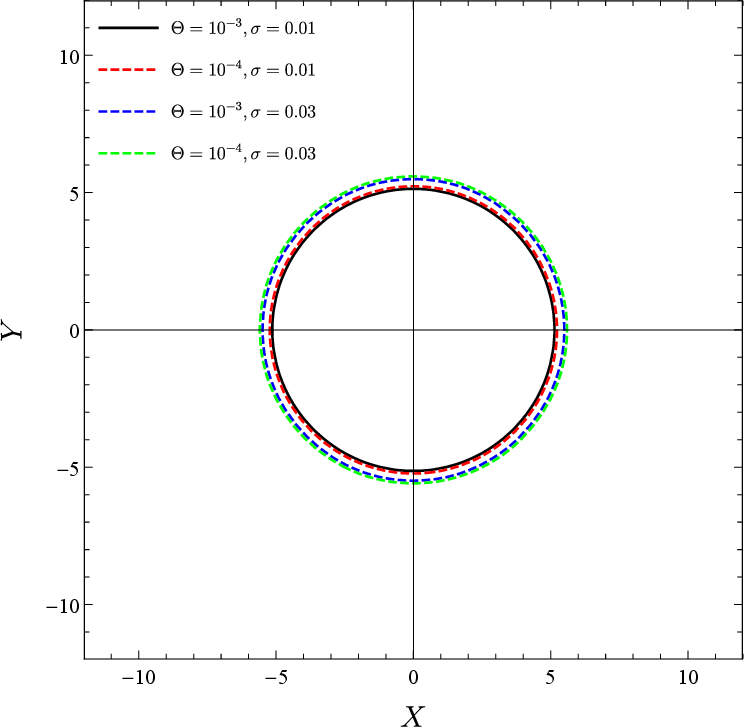}
         \subcaption{ $\omega_{q}=-0.55$.}\label{fig:shb}
   \end{minipage}\\ 
\begin{minipage}[t]{0.50\textwidth}
       \centering
        \includegraphics[width=\textwidth]{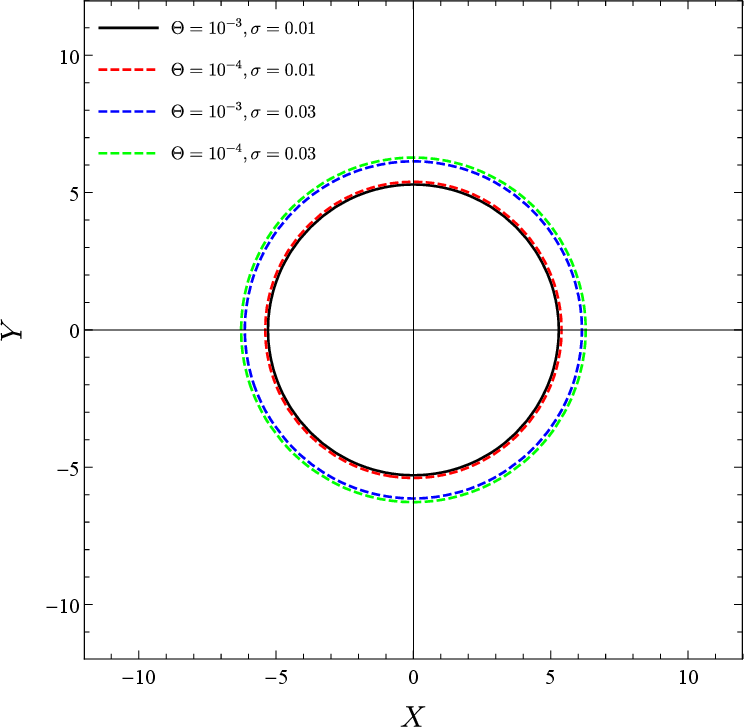}
       \subcaption{  $\omega_{q}=-0.75$.}\label{fip:shc}
   \end{minipage}%
\begin{minipage}[t]{0.50\textwidth}
        \centering
        \includegraphics[width=\textwidth]{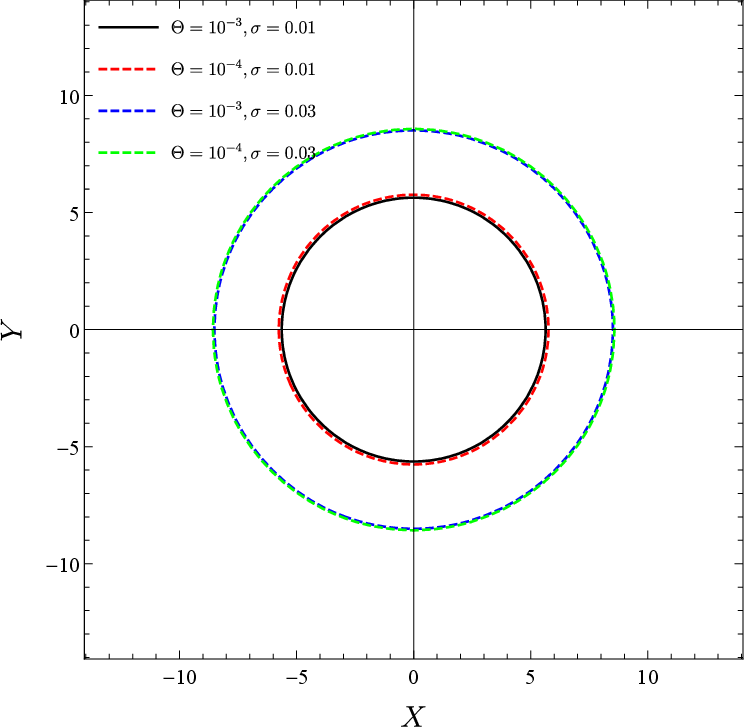}
       \subcaption{ $\omega_{q}=-0.95$.}\label{fip:shd}
   \end{minipage}
\caption{Black hole shadow in the celestial plane ($X$-$Y$) for different values of $\Theta $ and $\protect\sigma $ with $k=0.1$.}
\label{figsh2}
\end{figure}

\newpage
{ Figure \eqref{fig:pla} shows us that for a constant value $\Theta$ and $\sigma$ we should observe a larger shadow radius at smaller quintessence state parameter values. Moreover, Figure \eqref{fig:plb} tells us that for higher $\sigma$ values the shadow radius grows significantly. However, Figure \eqref{figsh1} cannot reveal the effect of noncommutative spacetime. To shed light on the noncommutativity effect, we show black hole shadows in Figure \eqref{figsh2} for four different constant quintessence state parameters. }
\newpage
{ We observe that for smaller noncommutativity parameters the shadow radius become slightly greater. We see that this effect remains the same at smaller quintessence state parameters. }

\newpage
\section{Quasinormal modes}\label{sec5}

In general, the QNMs can be derived within two methodologies: perturbing the metric \cite{Jurica, Juricb}, or coupling fields to the spacetime and considering the interactions between the black hole and the fields \cite{Mod1, Cho2003, Jing2004, Mod4, Mod3}. In this manuscript, we will follow the second methodology.

\subsection{Wentzel - Kramers - Brillouin approximation}

In this section, we explore the QNMs of the noncommutative Schwarzschild black hole surrounded by quintessence for a scalar field. We assume that the scalar field's influence on the black hole spacetime is insignificant, implying a negligible back reaction. Our approach to examining QNMs involves initially examining the Klein-Gordon equation and subsequently transforming
it into a Schr\"{o}dinger-like equation form. In the case of a massless scalar field, we {first} write the Klein-Gordon equation 
\begin{equation}
\frac{1}{\sqrt{-g}}\partial _{\mu }\left( \sqrt{-g}g^{\mu \nu }\partial
_{\nu }\psi \right) =0,  \label{klei}
\end{equation}
{we then} employ the separating variables method with the following assumption: 
\begin{equation}
\psi \left( t,r,\theta ,\varphi \right) =e^{-i\omega t}\frac{\Psi _{\omega
,L}\left( r\right) }{r}Y_{L,\mu }\left( \theta ,\varphi \right) ,
\label{dec}
\end{equation}%
where $\omega $ is the frequency, and $Y_{L,\mu }\left( \theta ,\varphi\right) $ are the spherical harmonics. By inserting {the described decomposition} into Eq. (\ref{klei}), we {get} a Schr\"{o}dinger like equation:%
\begin{equation}
\frac{d^2}{dr^{\ast ^2 }}\Psi \left( r^{\ast }\right) -\left( \omega ^{2}-%
\mathcal{V}\left( r\right) \right) \Psi \left( r^{\ast }\right) =0,
\label{eqto}
\end{equation}
with the tortoise variable, $dr^{\ast }=\frac{dr}{f\left( r\right) },$ and 
\begin{equation}
\mathcal{V}\left( r\right) =\frac{f\left( r\right) }{r}\frac{df\left(
r\right) }{dr}+\frac{f\left( r\right) L\left( L-1\right) }{r}.
\end{equation}
To solve Eq. (\ref{eqto}), it is necessary to take {suitable boundary conditions} into account. In this scenario, acceptable solutions are those that are purely ingoing near the horizon:%
\begin{equation}
\Psi \simeq e^{\pm i\omega r^{\ast }},\qquad r^{\ast }\rightarrow
\pm \infty .
\end{equation}
Now, we employ the WKB approximation to calculate the QNMs. Before giving the results, we would like to mention that the pioneering work using the WKB approximation to evaluate the QNMs was done by Schutz and Will \cite{Schutz}. Later, others extended {the approximation} to higher orders \cite{Iyer, Konoplya, Opala}. Following \cite{Konoplya, Opala}, one can obtain the frequencies of QNMs
with the following formula: 
\begin{equation}
i\frac{\left( \omega -V_{0}\right) }{\sqrt{-2V_{0}"}}-\sum_{i=2}^{N}\Lambda
_{i}=n+\frac{1}{2}.  \label{qnmf}
\end{equation}
Here, $V_{0}$ and $V_{0}"$ denote effective potential's height, and the
second derivative with respect to the tortoise coordinate of the potential
at its maximum "$r_{0}^{\ast }$", respectively. Additionally, $\Lambda _{i}$
is a constant coefficient resulting from higher order WKB corrections, and $%
n=0,1,2,...$, is the overtone number. { It is worth noting that the} explicit expressions of $\Lambda
_{i}$ for higher orders are given in \cite{Konoplya, Opala}.

{ We then utilize Eq. (\ref{qnmf}) to compute quasinormal frequency values with} scalar field perturbations across different angular {momentum  and quintessence matter field parameter} by using Pad\'{e} averaged 6th order WKB approximation method. It is important to note that, according to the WKB formula, the best accuracy is achieved for $L>n$ \cite{Zhidenko}. Consequently, we exclusively examine scalar field functions that adhere to this condition, as they are associated with the low-lying QNMs{.} We tabulate our results in Table \ref{tab:week3}.
\begin{table}[htb]
\centering%
\begin{tabular}{|l|l|l|l|l|l|}
\hline\hline
\rowcolor{lightgray} $L$ & $n$ & $\omega _{q}=-0.35$ & $\omega _{q}=-0.55$ & 
$\omega _{q}=-0.75$ & $\omega _{q}=-0.95$ \\ \hline
1 & 0 & 0.268812 - 0.103954 i & 0.264652 - 0.101739 i & 0.256376 - 0.098084 i & 
0.248113 - 0.094897 i \\ \hline
2 & 0 & 0.466582 - 0.102567 i & 0.459427 - 0.100353 i & 0.445267 - 0.096690 i & 
0.416550 - 0.091167 i \\ 
& 1 & 0.434087 - 0.379566 i & 0.429369 - 0.370736 i & 0.420170 - 0.356835 i & 
0.408909 - 0.337193 i \\ \hline
3 & 0 & 0.658273 - 0.099810 i & 0.648278 - 0.097661 i & 0.628470 - 0.094104 i & 
0.588260 - 0.088741 i \\ 
& 1 & 0.669246 - 0.326342 i & 0.658983 - 0.318950 i & 0.639112 - 0.306561 i & 
0.600805 - 0.286269 i \\ 
& 2 & 0.743689 - 0.489463 i & 0.731238 - 0.479461 i & 0.707431 - 0.462567 i & 
0.660694 - 0.436178 i \\ \hline
4 & 0 & 0.850307 - 0.098189 i & 0.837430 - 0.096083 i & 0.811909 - 0.092595 i & 
0.760102 - 0.087343 i \\ 
& 1 & 0.858471 - 0.308954 i & 0.845306 - 0.302195 i & 0.819498 - 0.290890 i & 
0.768118 - 0.273022 i \\ 
& 2 & 0.899347 - 0.570692 i & 0.884810 - 0.556975 i & 0.858113 - 0.534023 i & 
0.808330 - 0.444512 i \\ 
& 3 & 0.977143 - 0.678626 i & 0.960660 - 0.664834 i & 0.929234 - 0.641516 i & 
0.867694 - 0.605089 i \\ \hline\hline
\end{tabular}%
\caption{The QNMs of the noncommutative Schwarzschild black hole surrounded by quintessence for the massless scalar perturbation with $M=1$, $\Theta=10^{-4}$ and $\protect\sigma =0.01$ using the Pade averaged 6th order WKB approximation method.}
\label{tab:week3}
\end{table}

\begin{figure}[!htb]
\begin{minipage}[t]{0.5\textwidth}
        \centering
        \includegraphics[width=\textwidth]{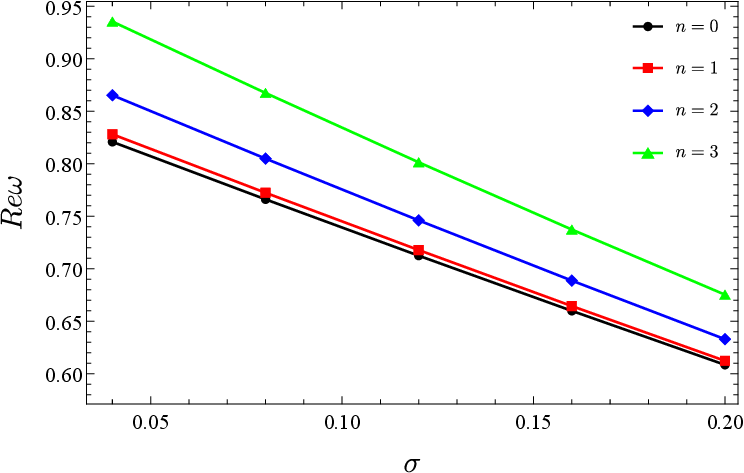}
       \subcaption{ $\omega_{q}=-0.35$.}\label{fig:qna}
   \end{minipage}%
\begin{minipage}[t]{0.5\textwidth}
        \centering
        \includegraphics[width=\textwidth]{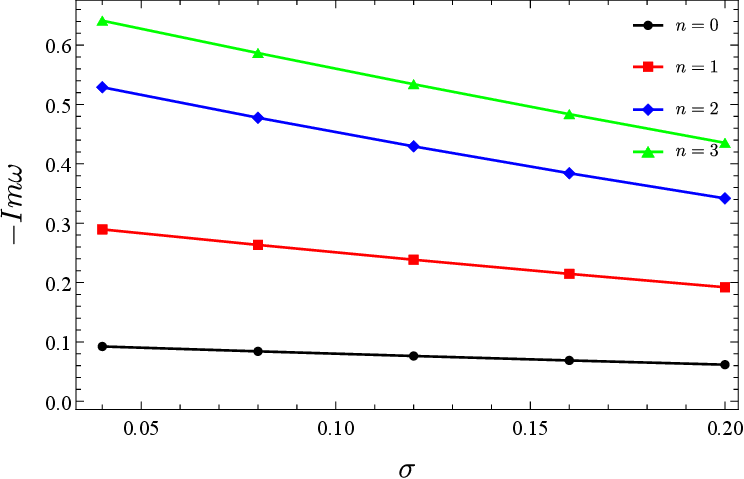}
         \subcaption{ $\omega_{q}=-0.35$.}\label{fig:qnb}
   \end{minipage}\ 
\begin{minipage}[t]{0.5\textwidth}
       \centering
        \includegraphics[width=\textwidth]{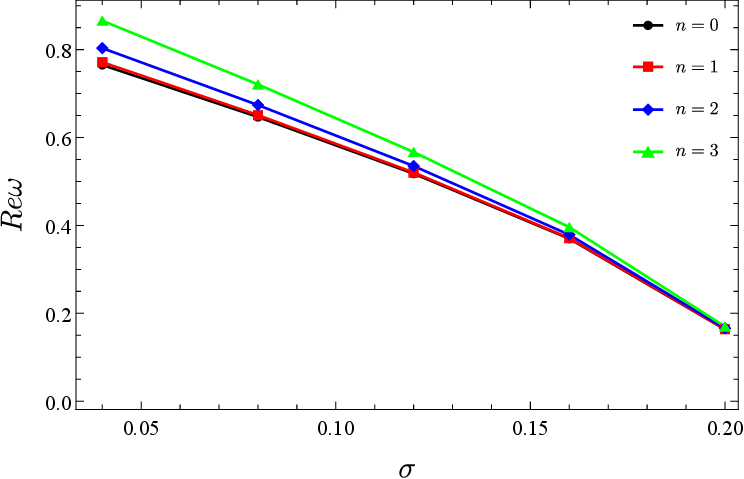}
       \subcaption{$\omega_{q}=-0.55$.}\label{fip:qnc}
   \end{minipage}%
\begin{minipage}[t]{0.5\textwidth}
        \centering
        \includegraphics[width=\textwidth]{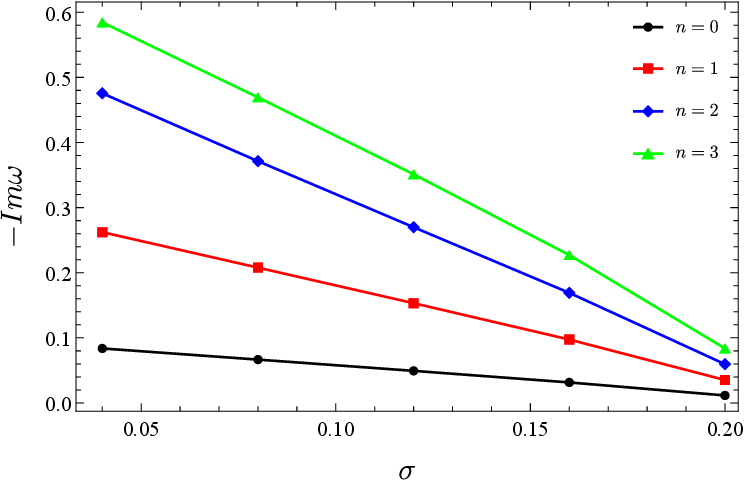}
       \subcaption{$\omega_{q}=-0.55$.}\label{fip:qnd}
   \end{minipage}
\caption{The variation of the $\text{Re}\,\protect\omega $ and $\text{Im}\,%
\protect\omega $ versus $\protect\sigma $ for two values of \ $\protect\omega%
_{q}=-0.35,-0.55$ with $\Theta =10^{-4}$. }
\label{Fig12}
\end{figure}

In Figure \ref{Fig12}, we compare the variation of  {real and imaginary components of the quasinormal frequencies versus $\sigma$, considering two different cases.} 

\begin{figure}[!htb]
\begin{minipage}[t]{0.5\textwidth}
        \centering
        \includegraphics[width=\textwidth]{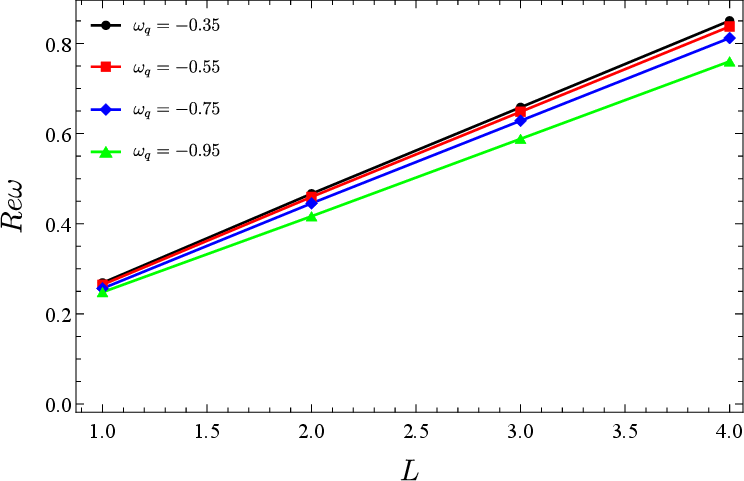}
       \label{fig:qnna}
   \end{minipage}%
\begin{minipage}[t]{0.5\textwidth}
        \centering
        \includegraphics[width=\textwidth]{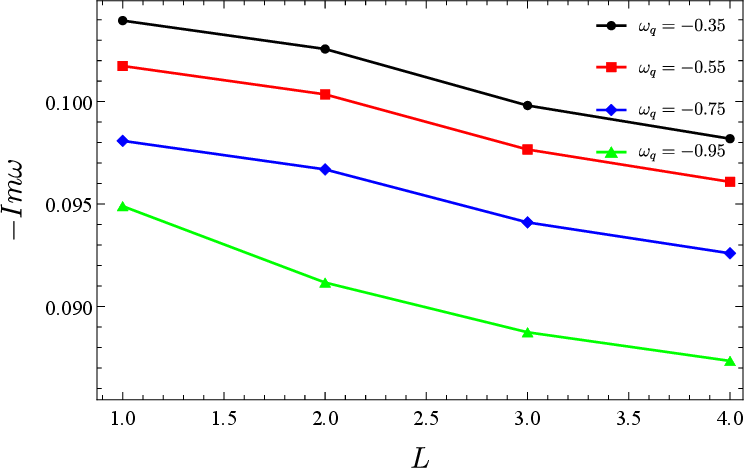}
        \label{fig:qnnb}
   \end{minipage}\ 
\caption{The variation of the \text{Re} $\omega $ and \text{Im} $\omega $ versus $L$ for $\sigma =0.01$, $n=0$ and $\Theta =
10^{-4}$. }
\label{qnmn}
\end{figure}

\begin{figure}[!htb]
\begin{minipage}[t]{0.5\textwidth}
        \centering
        \includegraphics[width=\textwidth]{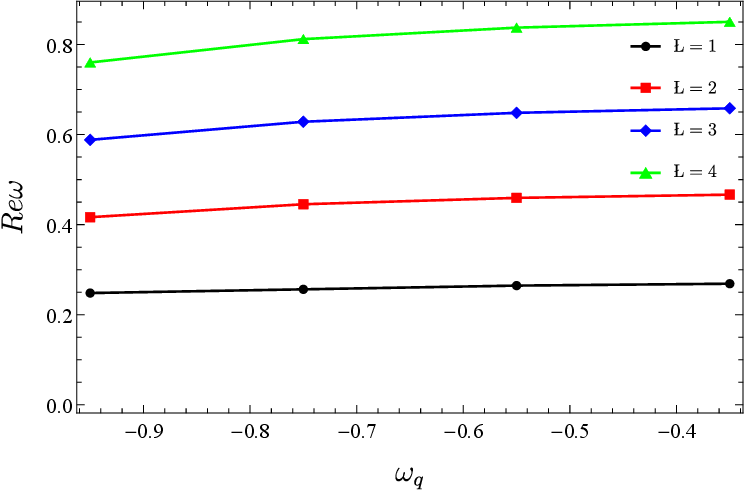}
       \label{fig:qnnna}
   \end{minipage}%
\begin{minipage}[t]{0.5\textwidth}
        \centering
        \includegraphics[width=\textwidth]{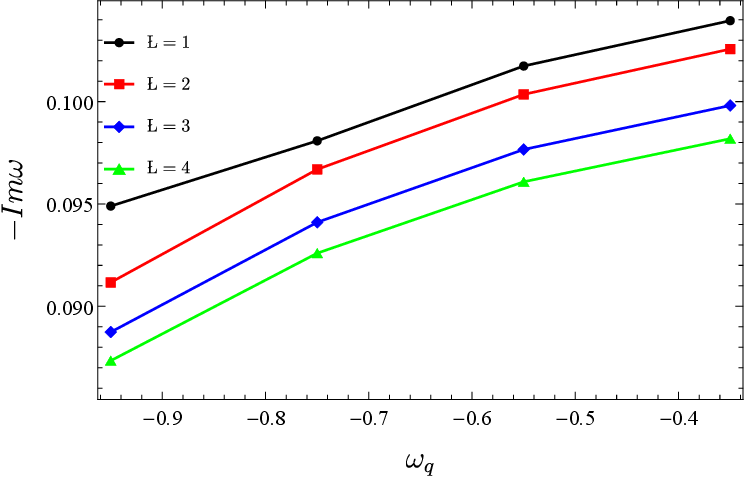}
        \label{fig:qnnnb}
   \end{minipage}\ 
\caption{The variation of the \text{Re} $\omega $ and  \text{Im} $\omega$ versus $\omega _{q}$ for $\sigma =0.01$, $n=0$ and $\Theta =10^{-4}$. }
\label{qnmL}
\end{figure}

We observe that with higher quintessence parameters both parts of
the frequencies decrease. This means that the presence of the quintessence matter damps the oscillations. Then, we repeat the a {a similar comparison based on the angular momentum} in Figure \ref{qnmn}, and the quintessence state parameter in Figure \ref{qnmL}, respectively. We conclude that the presence of the quintessence matter field alters the frequencies. 
\newpage

As we know, the WKB order plays a crucial role in obtaining the accurate values of QNMs frequencies. Increasing the order typically leads to improved approximations. However, this trend doesn't hold when we raise the multipole number. This observation becomes evident through the error estimatiom associated with each order of the WKB formula. This error quantity for $\omega _{k}$, derived from the WKB formula of order $k$ for each overtone $n$, is defined as:%
\begin{equation}
\Delta _{k}=\frac{\left\vert \omega _{k+1}-\omega _{k-1}\right\vert }{2}.
\end{equation}
Using the definition, we present the calculated error estimation in Table \ref{tabm3}.
\begin{table}[htb]
\centering%
\begin{tabular}{|l|ll|ll|}
\hline
\rowcolor{lightgray} \multirow{2}{*}{} & \multicolumn{2}{|c|}{$\omega
_{q}=-0.35$} & \multicolumn{2}{|c|}{$\omega _{q}=-0.55$} \\ \hline\hline
\rowcolor{lightgray} $k$ & $\omega _{k}$ & $\Delta _{k}$ & $\omega _{k}$ & $%
\Delta _{k}$ \\ \hline
$3$ & 0.820625 - 0.0911553 i & 0.00639599 & 0.781027 - 0.085026 i & 0.00584784
\\ 
$4$ & 0.821186 - 0.0960695 i & 0.00347966 & 0.781522 - 0.089458 i & 0.00311287
\\ 
$5$ & 0.826060 - 0.0955027 i  & 0.00433174 & 0.785848 - 0.088966 i & 0.00379935
\\ 
$6$ & 0.825272 - 0.0884303 i & 0.00753273 & 0.785173 - 0.082794 i & 0.00648148
\\ 
$7$ & 0.812089 - 0.0898658 i  & 0.01604200 & 0.773872 - 0.084003 i & 0.00648148
\\ 
$8$ & 0.815841 - 0.1190980 i & 0.03899340 & 0.776933 - 0.108640 i & 0.03252780 \\ 
$9$ & 0.887570 - 0.1094730 i  & 0.11296100 & 0.836701 - 0.100879 i & 0.09267570 \\ 
\hline
\rowcolor{lightgray} \multirow{2}{*}{} & \multicolumn{2}{|c|}{$\omega
_{q}=-0.75$} & \multicolumn{2}{|c|}{$\omega _{q}=-0.95$} \\ \hline\hline
$3$ & 0.696617 - 0.0743361 i & 0.00492351 & 0.490084 - 0.0538299 i & 0.00326404
\\ 
$4$ & 0.697008 - 0.0779159 i & 0.00246121 & 0.490306 - 0.0558108 i & 0.00125508
\\ 
$5$ & 0.700350 - 0.0775440 i & 0.00285470 & 0.491825 - 0.0556384 i & 0.00157382 \\ 
$6$ & 0.699858 - 0.0729691 i & 0.00465141 & 0.491631 - 0.0538938 i & 0.00116516
\\ 
$7$ & 0.691827 - 0.0738162 i & 0.00933632 & 0.489036 - 0.0541798 i & 0.00267457
\\ 
$8$ & 0.693824 - 0.0906397 i & 0.02164900 & 0.489573 - 0.0588313 i & 0.00536645 \\ 
$9$ & 0.733450 - 0.0857426 i & 0.05924060 & 0.499175 - 0.0576996 i & 0.01252900 \\ 
\hline\hline
\end{tabular}%
\caption{QNMs of the massless scalar field for $M=1$, $\protect\sigma =0.03$%
, $\Theta =10^{-4}$ and $L=4$, $n=0$ calculated with the WKB formula of
different orders.}
\label{tabm3}
\end{table}
\newpage

\subsection{Mashhoon approximation}

The Mashhoon method \cite{Mashhoon}, involves the approximation of the potential $\mathcal{V}\left( r^{\ast }\right) $ by using a potential with a comparable shape, for which the solutions of equation (\ref{eqto}) can be evaluated analytically. This condition is met by employing the P\"{o}schl-Teller (PT) potential, which has the form 
\begin{equation}
\mathcal{V}\left( r^{\ast }\right) \sim V_{PT}\left( r^{\ast }\right) =\frac{%
V_{0}}{\cosh ^{2}\alpha \left( r^{\ast }-r_{0}^{\ast }\right) }.
\end{equation}%
Here, $\alpha $ is the curvature of the potential $\mathcal{V}\left( r^{\ast
}\right) $ at its maximum $r^{\ast }=r_{0}^{\ast }.$ Thus 
\begin{equation}
\ \alpha =\left. \sqrt{\frac{-1}{2V_{0}}\frac{d^{2}\mathcal{V}\left( r^{\ast
}\right) }{dr^{\ast 2}}}\right\vert _{r^{\ast }=r_{0}^{\ast }}.
\end{equation}
Solving Eq. (\ref{eqto}) with the PT approximation leads to the following
equation for the calculation of QNMs \cite{Mashhoon1}%
\begin{equation}
\omega =\pm \alpha \sqrt{\frac{V_{0}}{\alpha ^{2}}-\frac{1}{4}}+i\alpha
\left( n+\frac{1}{2}\right) .  \label{Mas}
\end{equation}
We compute the QNMs using the Mashhoon approximation via Eq. (\ref{Mas}). We tabulate the results in Table \ref{tabComparison}, and subsequently we compare them with ones obtained through the WKB method.
\begin{table}[htbp]
\centering%
\begin{tabular}{|l|ll|ll|}
\hline
\rowcolor{lightgray} \multirow{2}{*}{} & \multicolumn{2}{|c|}{$\omega
_{q}=-0.35$} & \multicolumn{2}{|c|}{$\omega _{q}=-0.90$} \\ \hline\hline
\rowcolor{lightgray} $\Theta\times
10^{-4} $ & WKB method & Mashhoon appr. & WKB method & 
Mashhoon appr. \\ \hline
$1$ & 0.289890-0.117280 i & 0.275811-0.0902164 i & 0.0602218-0.0244681 i & 
0.0574887-0.0200356 i \\ 
$2$ & 0.290632-0.117272 i & 0.276638-0.0902846 i & 0.0634264-0.0257514 i & 
0.0605687-0.0210856 i \\ 
$4$ & 0.291696-0.117255 i & 0.277826-0.0903789 i & 0.0678057-0.0274948 i & 
0.0647827-0.0225133 i \\ 
$6$ & 0.292526-0.117237 i & 0.278753-0.0904494 i & 0.0710663-0.0287839 i & 
0.0679237-0.0235701 i \\ 
$8$ & 0.293233-0.117218 i & 0.279545-0.0905075 i & 0.0737591-0.0298420 i & 
0.0705202-0.0244384 i \\ 
$10$ & 0.293864-0.117198 i & 0.280252-0.0905576 i & 0.0760948-0.0307546 i & 
0.0727743-0.0251882 i \\ \hline\hline
\end{tabular}%
\caption{Comparison of QNMs for a scalar field of noncommutative Schwarzschild black hole surrounded by quintessence, derived through the Mashhoon approximation and { 3rd order} WKB method, for specific parameters: $M=1$, $L=1$, $\sigma =0.05$, and varying values of the parameter $\Theta $.}
\label{tabComparison}
\end{table}
\normalsize

\section{ Conclusion}

In this work, we intended to investigate thermodynamics, shadows, and QNMs
features of the Schwarzschild black hole surrounded by the quintessence
matter in noncommutative spacetime. To achieve our goal, we first introduced
the lapse function of the black hole, and then, we found the smeared mass
function in terms of the event horizon. In four different scenarios of
quintessence matter fields, we demonstrated the mass function's
characteristics. We found that the event horizon has an upper bounded value
depending on the quintessence matter field. Next, we obtained the Hawking
temperature and verified our result with the ones that exist in the
literature by considering the limit values of the scenarios. We noticed that
the noncommutativity eliminates the divergence problem of the Hawking
temperature. In addition, we noted that the black hole temperature rises
during its evaporation, and it reaches a peak value $T_{H}^{\max }$ at a
critical horizon radius value, and subsequently, it decreases to zero
rapidly. Then, we derived the entropy function and observed that the
quintessence matter does not alter its form, however, noncommutative effects
do with a positive valued contribution. Next, we studied the heat capacity
function. Its complex form allowed us to discuss its characteristics only
numerically. We observed that the quintessence state parameter value has a
critical role in black hole stability. In some scenarios, we found that the
black hole can only be in the unstable state, but in other scenarios, we
noted both the unstable and stable states of the black hole. We have also
determined that the black hole radiation will terminate and a remnant mass
will form. To understand whether the stability is local or global, we
considered the Gibbs function and noted that the turning point varies
depending on both the quintessence matter and noncommutative effects. Next,
we studied the shadows by considering plasma distribution. After we obtained
the effective potential, we showed the impact of the quintessence matter on
the potential and shadows. Moreover, following the numerical calculations,
we tabulated the photon radius and impact parameters. In the final section,
we studied QNMs with WKB approximation. We observed the damping effects of
the quintessence matter on reel and imaginary parts of frequencies. After
visualizing our results, we compared the QNMs modes with those obtained with
the Mashhoon approach.

\section*{Acknowledgments}
{ The authors are thankful to the anonymous reviewers for their constructive comments.} B. C. L. is grateful to Excellence project PřF UHK 2211/2023-2024 for the financial support.

\section*{Data Availability Statements}

The authors declare that the data supporting the findings of this study are
available within the article.

\end{document}